\documentclass[reprint,superscriptaddress,nofootinbib,amsmath,amssymb,prd,floatfix]{revtex4-1}
\pdfoutput=1
\usepackage{graphicx}
\usepackage[dvipsnames]{xcolor}
\usepackage{amsmath}
\allowdisplaybreaks
\usepackage{amssymb}

\usepackage{slashed}
\usepackage[utf8]{inputenc}
\usepackage[colorlinks=true,linkcolor=blue,bookmarksopen,bookmarksnumbered]{hyperref}
\usepackage{color}
\usepackage[normalem]{ulem}
\usepackage{soul}
\usepackage{units}
\usepackage{rotating}
\usepackage{hhline,multirow,tabularx}
\usepackage{bm}
\usepackage{hyperref}
\usepackage[export]{adjustbox}
\usepackage{mdframed}
\usepackage{comment}
\usepackage{natbib}

\def\bea#1\eea{\begin{align}#1\end{align}} 

\newcommand{\bef}{\begin{figure}[htb]\centering}
\newcommand{\eef}{\end{figure}}

\newcommand{\nn}{\nonumber}

\def\L{\Lambda}

\def\<{\langle}
\def\>{\rangle}

   \def\L{\Lambda}

\def\({\left(}
\def\[{\left[}
\def\){\right)}
\def\]{\right]}

\def\sin{\hbox{sin}}
\def\ln{\hbox{ln}}

\usepackage{lipsum}
\begin{document}

\title{Transverse $\L$ Polarization in $e^+ e^-$ collisions}
	
\author{Leonard Gamberg}
\email{lpg10@psu.edu}
\affiliation{Division of Science$,$ Penn State Berks$,$ Reading$,$ Pennsylvania 19610$,$ USA}
	
\author{Zhong-Bo Kang}
\email{zkang@g.ucla.edu}
\affiliation{Department of Physics and Astronomy, University of California, Los Angeles, California 90095, USA}
\affiliation{Mani L. Bhaumik Institute for Theoretical Physics, University of California, Los Angeles, California 90095, USA}
\affiliation{Center for Frontiers in Nuclear Science, Stony Brook University, Stony Brook, New York 11794, USA}

\author{Ding Yu Shao}
\email{dingyu.shao@cern.ch}
\affiliation{Department of Physics and Astronomy, University of California, Los Angeles, California 90095, USA}
\affiliation{Mani L. Bhaumik Institute for Theoretical Physics, University of California, Los Angeles, California 90095, USA}
\affiliation{Center for Frontiers in Nuclear Science, Stony Brook University, Stony Brook, New York 11794, USA}

\author{John Terry}
\email{johndterry@physics.ucla.edu}
\affiliation{Department of Physics and Astronomy, University of California, Los Angeles, California 90095, USA}
\affiliation{Mani L. Bhaumik Institute for Theoretical Physics, University of California, Los Angeles, California 90095, USA}

\author{Fanyi Zhao}
\email{fanyizhao@physics.ucla.edu}
\affiliation{Department of Physics and Astronomy, University of California, Los Angeles, California 90095, USA}
\affiliation{Mani L. Bhaumik Institute for Theoretical Physics, University of California, Los Angeles, California 90095, USA}


\begin{abstract}
In this paper we study transverse polarization of $\L$ hyperons in single-inclusive leptonic annihilation. We show that when the transverse momentum of the $\L$ baryon is measured with respect to the thrust axis, a transverse momentum dependent (TMD) factorization formalism is required and the polarization is generated by the TMD polarizing fragmentation function (TMD PFF), $D_{1T}^\perp$. However, when the transverse momentum of the $\L$ baryon is measured with respect to the momentum of the initial leptons, a collinear twist-3 formalism is required and the polarization is generated by the intrinsic collinear twist-3 fragmentation function $D_{T}$. Thus while these measurements differ from one another only by a change in the measurement axis, they probe different distribution functions. Recently, Belle measured a significant polarization in single-inclusive $\L$ baryon production as a function of the transverse momentum with respect to the thrust axis. However, this data can in principle be re-analyzed to measure the polarization as a function of the transverse momentum of the $\L$ baryon with respect to the lepton pair. This observable could be the first significant probe of the function, $D_{T}$. In this paper, we first develop a TMD formalism for $\L$ polarization; we then present a recent twist-3 formalism that was established to describe $\L$ polarization. Using the TMD formalism, we demonstrate that the $\L$ polarization at OPAL and Belle can be described using the twist-2 TMD factorization formalism. Finally, we make a theoretical prediction for this polarization in the collinear twist-3 formalism at Belle.
\end{abstract}
\maketitle
\section{Introduction}\label{Introduction}

It has been a long standing challenge to describe the transverse polarization of $\L$ baryons in deep inelastic high energy reactions from a factorized framework in perturbative QCD. The strikingly large transverse polarization asymmetries  of $\L$ hyperons  observed in early experiments at Fermilab (along with follow-up experiments) in $pA\to \L X$ fixed target processes already 40 years ago  \cite{Bunce:1976yb,Schachinger:1978qs,Heller:1983ia}, was at odds with the predictions from transverse polarization effects in perturbative QCD~\cite{Kane:1978nd}. The discrepancy between theory and experiment has resulted in numerous 
experimental~\cite{,Lundberg:1989hw,Yuldashev:1990az,Ramberg:1994tk} and theoretical investigations~\cite{Kane:1978nd,Panagiotou:1989sv,Dharmaratna:1996xd,Anselmino:2001la,Anselmino:2001js,Boer:2010ya,Boer:2010yp,Wei:2014pma,Gamberg:2018fwy} that have spanned decades.
Fixed target measurements of this reaction were reported by the NA48 collaboration~\cite{Fanti:1998px} and the HERA-B collaboration~\cite{Abt:2006da}. At CERN the $\L$ polarization was also measured in $pp$ collisions at moderate center-of-mass (CM) energy~\cite{Erhan:1979xm}. More recently, polarization of $\L$ baryons were investigated at the LHC by the ATLAS collaboration \cite{ATLAS:2014ona}. While a small polarization was found in the ATLAS results in the mid-rapidity region measurements, essentially consistent with zero, such experiments demonstrate that the polarization of $\L$ baryons can be studied at the highest LHC energies and may be larger in different kinematical regions at forward rapidities. 

Experimentally, data on polarized $\L$ fragmentation in $e^+e^-$-annihilation  has been provided by the OPAL collaboration \cite{Ackerstaff:1997nh} at the LEP. This measurement was performed on the $Z$-pole, i.e., at a center of mass energy equal to the mass of the $Z$-boson. While a substantial  {\it longitudinal} polarization of the $\L$s was detected by OPAL, the {\it transverse} polarization was found to be zero within error bars.

Recently the Belle collaboration   measured the production of transverse polarization of $\L$-hyperons~\cite{Guan:2018ckx} in $e^+e^-$-annihilation for single-inclusive $\L$ production, where the hadron cross section is studied as a function of the fractional energy $z_{\L}$, and the transverse momentum $\bm j_\perp$ with respect to the thrust axis. They find a significant non-zero effect for this process as well as for back-to-back production of $\L$ and a light hadron $h=\pi^\pm,~ K^\pm$. 

From theory there has been much progress since the work in Ref.~\cite{Kane:1978nd}. For processes  with more than one hard scale, such as the case for $\L$ production in semi-inclusive deep inelastic scattering (SIDIS) as well as back-to-back $\L + h$   production in $e^+ e^-$
collisions in the Belle experiment, the transverse momentum dependent (TMD) formalism predicts a non-trivial result in term of TMD fragmentation functions (FFs)~\cite{Mulders:1995dh}. In the TMD factorization  framework~\cite{Collins:1981uk,Boer:1997mf,Collins:2011zzd} for back-to-back production of $\L + h$,  a chiral even, naively $T$-odd fragmentation function, the TMD polarizing fragmentation function (TMD PFF) $D_{1T}^\perp(z_\L, p_\perp)$ is predicted to be non-zero and universal~\cite{Collins:1992kk,Metz:2002iz,Collins:2004nx,Meissner:2008yf,Boer:2010ya,Gamberg:2010uw}. As a result of this Belle measurement, first phenomenological extractions of the T-odd polarizing TMD $D_{1T}^\perp$ were carried out recently in~\cite{DAlesio:2020wjq,Callos:2020qtu,Chen:2021hdn}. 

\begin{figure}[h]
    \includegraphics[width = 0.23\textwidth]{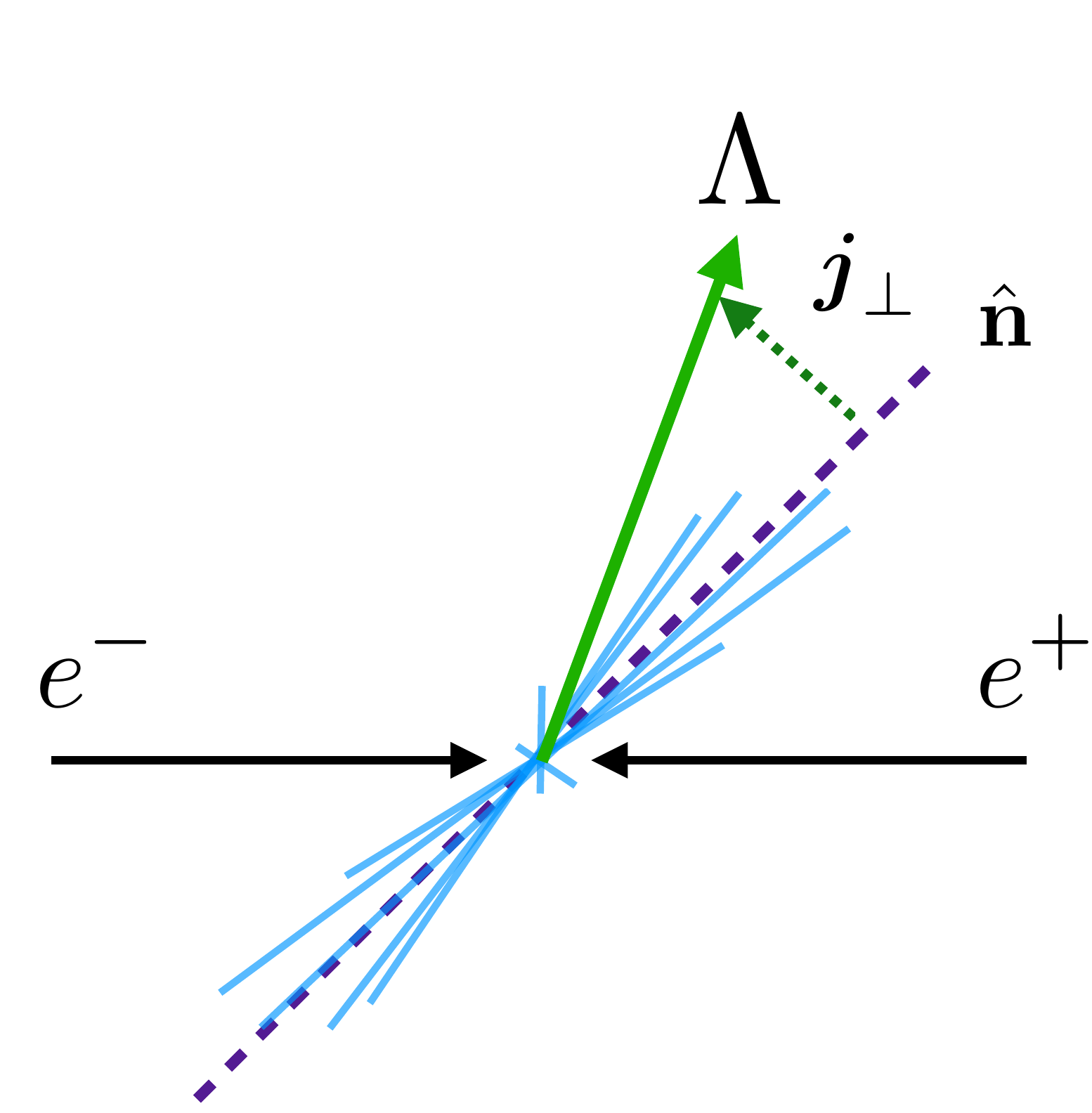}
    \hfill
    \includegraphics[width = 0.23\textwidth]{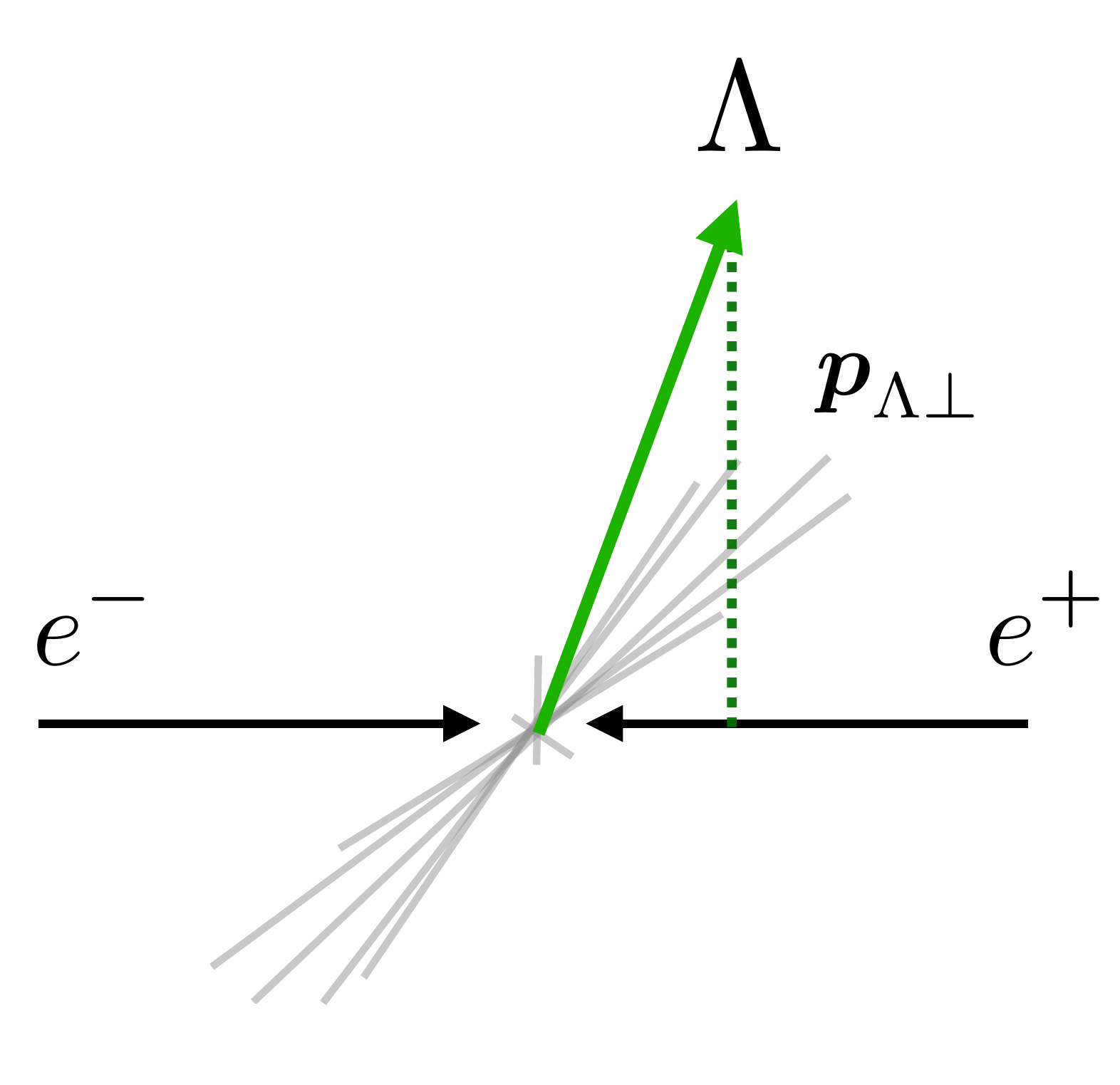}
 \caption{Left: Thrust reference frame $e^+e^-\rightarrow \L\text{(Thrust)} +X$. Right: Center-of-mass frame $e^+e^-\rightarrow \L +X$. }
        \label{fig:cartoons}
\end{figure}

While TMD factorization theorems have been well established for back-to-back production of $\L + h$~\cite{Collins:1981uk,Collins:1981uw,Collins:1981va,Collins:2011zzd}, the factorization for the thrust-axis process with unpolarized hadron production has only recently been considered from theory~\cite{Kang:2020yqw,Boglione:2020auc,Makris:2020ltr} in a TMD framework. In this case for $e^+e^-\rightarrow \L {\rm (Thrust}{\rm )} X$, as shown in Fig.~\ref{fig:cartoons} (left), one measures $\L$ transverse momentum $\bm j_\perp$ with respect to the thrust axis $\hat {\textbf{n}}$. Here, we extend this TMD factorization formalism to describe transversely polarized $\L$ production in this case with full QCD evolution. Establishing such a factorization  theorem is an essential tool to carry out a global analysis of the TMD PFF.

On the other hand, much of the above mentioned data have been for single inclusive $\L$ production, $e^+e^-\rightarrow \L\, X$, where there is a single hard scale -- the transverse  momentum $\bm p_{\L\perp}$ of the~$\L$, measured in the lepton center-of-mass (CM) frame as shown in Fig.~\ref{fig:cartoons} (right). In recent years QCD collinear factorization at higher twist~\cite{Qiu:1998ia,Metz:2012ct} predicts a non-trivial result for these processes. For fully inclusive $e^+e^-\rightarrow \L\, X$ the collinear twist-3 factorization framework predicts~\cite{Gamberg:2018fwy},  that the cross section factorizes into a hard scattering contribution and the collinear twist-3 polarizing fragmentation function, $D_{T}(z_\L)$. A treatment of transverse polarization for this process was also given in terms of a power suppressed, one particle inclusive cross section by Boer et al.~\cite{Boer:1997mf}, and was also studied earlier for the inclusive deep inelastic scattering (DIS) process~\cite{Lu:1995rp}.
It is interesting to note that by naive time reversal in what is expected to be the dominant  one photon production approximation $D_{T}(z_\L)$,    is predicted  non-zero~\cite{DeRujula:1972te, Collins:1992kk,Goeke:2005hb,Metz:2016swz}.

It is quite interesting that while these two measurements probe different distribution functions, they differ only by the definition of the measurement axis. That is, a measurement of the polarization as a function of $\bm j_\perp$ with respect to the thrust axis is a useful process for probing  the properties of the TMD PFF  $D_{1T}^\perp$, while  a measurement of the polarization as a function of $\bm p_{\L\perp}$,  the transverse momentum of the $\L$ in the lepton CM frame, is a useful process for probing the collinear twist-3 function,  $D_{T}$. 
Therefore the polarization in the CM frame can in principle be studied from the existing Belle data by re-analyzing the data for the inclusive $e^+e^-\rightarrow \L {\rm (Thrust}{\rm )} X$ measurement. With regard to the latter measurement, it is important to note that an observation of a non-zero effect  in the single  inclusive process, is a fundamental test of naive time reversal invariance~\cite{Christ:1966zz,DeRujula:1972te,Collins:1992kk,Lu:1995rp} which predicts a non-zero result for T-odd fragmentation, and a zero result for inclusive DIS processes~\cite{Goeke:2005hb}.
Furthermore, in the recent paper~\cite{Gamberg:2018fwy} the factorization of this process has been studied  at next to leading order in perturbative QCD. In this paper, we use this formalism to make a theoretical prediction at Belle for this process. In this paper, we provide a clear distinction between the TMD and twist-3 factorization theorems for these two measurements and in turn.

Our paper is organized as follows: In Sec.~\ref{TMD-Theory}, we provide the theoretical formalism for the $e^+\, e^-\rightarrow \, \L\, ({\rm Thrust})\, X$ process. In Sec.~\ref{Twist3-Theory}, we provide the theoretical formalism for the $e^+\, e^-\rightarrow \, \L\, X$ process. In Sec.~\ref{TMD-Pheno}, we provide the details of our phenomenological analysis for the thrust TMD formalism and make a comparison of our formalism against the measurements performed by OPAL and Belle. In Sec.~\ref{Twist-Pheno}, we provide a theoretical prediction at Belle kinematics. We conclude our paper in Sec.~\ref{Conclusion}. 

\section{QCD factorization}\label{theory}
In this section, we provide the theoretical framework of our analysis. In Sec.~\ref{TMD-Theory}, we extend the theoretical formalism presented in \cite{Kang:2020yqw} to describe transverse polarization in $e^+\, e^-\rightarrow \, \L\, (\rm{Thrust})\, X$ as shown in the left side of Fig.~\ref{fig:cartoons}, where $\bm j_\perp$ is the $\L$ transverse momentum with respect to the thrust axis $\hat{\textbf{n}}$. In Sec.~\ref{Twist3-Theory}, we provide the formalism for transverse $\L$ polarization in the twist-3 collinear formalism under center-of-mass frame as illustrated in the right side of Fig.~\ref{fig:cartoons}, where $\bm p_{\L \perp}$ is the transverse momentum of the $\L$ baryon relative to the momentum of incoming electron. 

\subsection{$\L$ Polarization in the Thrust Frame}\label{TMD-Theory}
\begin{figure}[hbt!]
    \centering
    \includegraphics[width = 0.45\textwidth]{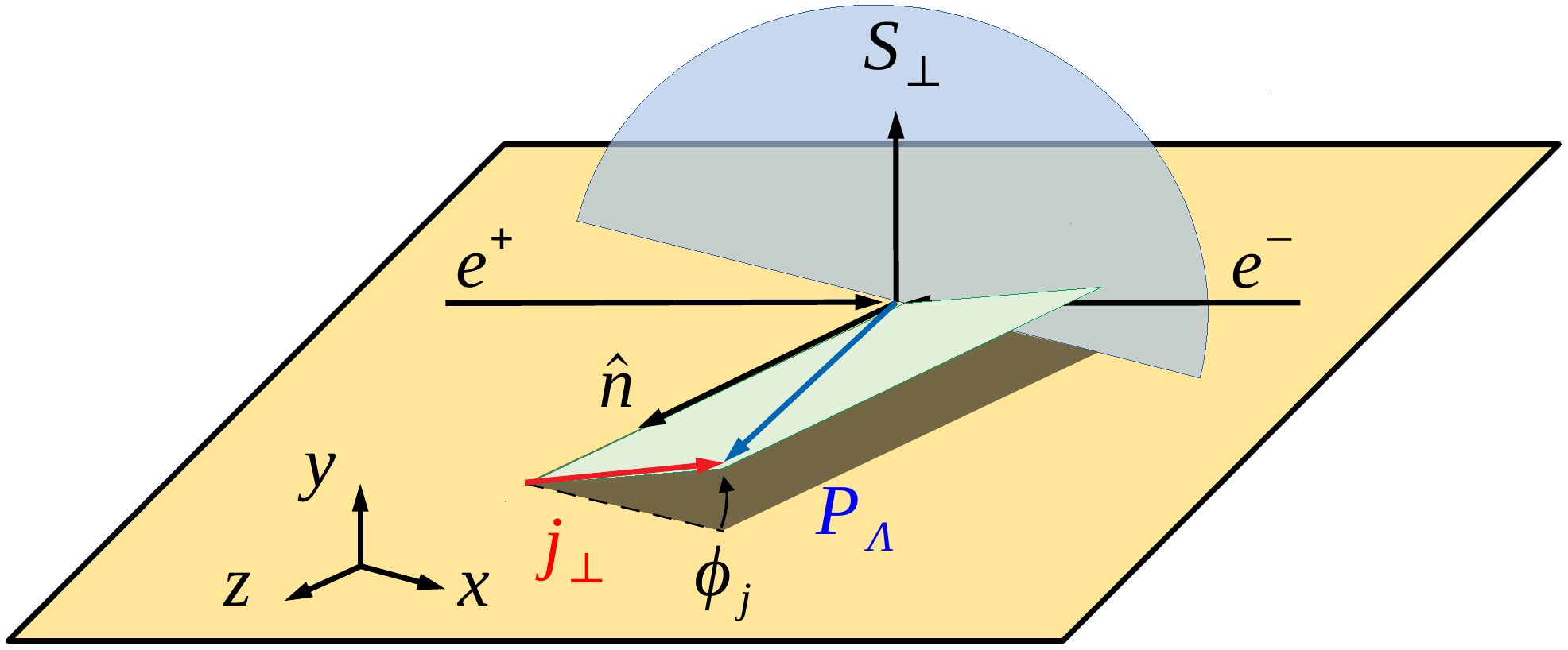}
    \caption{Transverse $\L$ polarization in the thrust frame. The blue semi-circle represents the plane which is perpendicular to the thrust axis $\hat{\textbf{n}}$. }
    \label{fig:thrust}
\end{figure}
In this section, we consider the transverse polarization for the process
\begin{align}
    e^-(l) +e^+(l') \rightarrow \gamma^*(q) \rightarrow \L\big(z_\L, \bm{j}_{\perp}, \bm{S}_{\perp}\big) + X\,.
\end{align}
In this expression, $q^\mu=l^\mu+l'^\mu$ with $Q \equiv \sqrt{q^2}$, and $z_\L = 2 P_\L\cdot q/Q^2$ is the parton fraction variable for the fragmentation function while the center-of-mass energy for this process is given by $s = Q^2$. The momentum $\bm{j}_{\perp}$ represents the transverse momentum of the $\L$ baryon with respect to the thrust axis, $\hat{\textbf{n}}$. The thrust axis is defined as the vector, $\hat{\textbf{n}}$, which maximizes the thrust variable $T$
\begin{align}
    T = \frac{\sum_i \left| \bm{p}_i\cdot \hat{\textbf{n}} \right|}{\sum _i \left| \bm{p}_i \right|}\,,
\end{align}
where $\bm{p}_i$ represent the momentum of the measured particles in the $e^+e^-$ collision. The plane which lies perpendicular to the thrust axis at the interaction point of the lepton pair divides the full phase space into two hemispheres. This plane is illustrated in Fig.~\ref{fig:thrust} by the blue semi-circular plane. Finally $\bm{S}_{\perp}$ is the transverse spin of the $\L$ baryon. 

In this paper, we consider the TMD kinematic region, i.e.~$j_\perp \ll  Q$. In this kinematic region, the factorized expression is given at next-to-leading logarithmic level (NLL) by~\cite{Kang:2020yqw}
\begin{align} \label{eq:unp-nll}
    \frac{d\sigma}{d z_{\L} d^2 \bm{j}_\perp}   =&\, \sigma_0 \sum_{q} e_q^2 \int d^2 p_\perp d^2\lambda_\perp\\
    & \times \delta^{(2)}\big(\bm{j}_\perp-\bm{p}_\perp-z_\L \bm{\lambda}_\perp \big) \nn \\
    & \times  D_{ \L/q}(z_{\L},p_\perp, \mu,\zeta/\nu^2) S_{\rm hemi}(\lambda_\perp,\mu,\nu) \nn \,,
\end{align}
where  
\begin{align}
\sigma_0 = \frac{4N_c \pi\alpha_{\rm em}^2}{3Q^2}\,.
\end{align} 
In this expression, $\mu$ is the renormalization scale, $\nu$ is the scale which is used to regulate the rapidity divergences~\cite{Chiu:2011qc,Chiu:2012ir}, and $\zeta$ is the Collins-Soper parameter~\cite{Ebert:2019okf,Collins:2011zzd}. We have also introduced $S_{\rm hemi}(\lambda_\perp,\mu,\nu)$, the hemisphere soft function, and $D_{\L/q}(z_\L,p_\perp,\mu,\zeta/\nu^2)$, the unpolarized TMD FF. It is important to emphasize that the hemisphere soft function is different than the usual soft function $S$ defined in \cite{Collins:2011zzd}, which is used to describe the back-to-back di-hardon production in $e^+e^-$ collisions. This difference occurs because $S_{\rm hemi}(\lambda_\perp,\mu,\nu)$ includes radiation in a single hemisphere while $S$ includes radiation in both hemispheres.

Furthermore, we note that the factorization theorem in Eq.~\eqref{eq:unp-nll} is a simplification of the full formula in~\cite{Kang:2020yqw}. This factorization matches the full one at NLL, while a more complicated factorization formula occurs at higher order. For example, the one-loop hard function in the full formula contains not only virtual corrections but also the wide-angle energetic radiation in one hemisphere. As a result, the gluon TMD FF also contributes to the factorized cross section at the leading power of $\mathcal{O}(\bm j_\perp^{\,2}/Q^2)$. For more details, see Ref.~\cite{Kang:2020yqw}.

TMD factorization is conventionally carried out in $b$-space, where the factorized expression deconvolutes~\cite{Boer:2011xd},
\begin{align}
    \frac{d \sigma}{d z_{\L} d^2 \bm{j}_\perp} & =\, \sigma_0 \sum_{q} e_q^2 \int_0^\infty \frac{d^2b}{(2\pi)^2} e^{i\bm{b}\cdot \bm{j}_\perp/z_\L}  \nn \\
    & \hspace{0.35cm} \times D_{ \L/q}(z_{\L},b, \mu,\zeta/\nu^2) S_{\rm hemi}(b,\mu,\nu) \,.
    \label{eq:unp-nll-FT}
\end{align}
In this expression, 
\begin{align}
    D_{ \L/q}(z_{\L},b,\mu,\zeta/\nu^2)  = & \frac{1}{z_\L^2} \int d^2p_\perp e^{-i \bm{b}\cdot \bm{p}_\perp /z_\L} \nonumber\\
    & \times D_{ \L/q}(z_\L,p_\perp,\mu,,\zeta/\nu^2)\,,
\\
    S_{\rm hemi}(b,\mu,\nu) =& \int d^2\lambda_\perp e^{-i \bm{b}\cdot \bm{\lambda}_\perp }S_{\rm hemi}(\lambda_\perp,\mu,\nu)\,,
\end{align}
are the Fourier transforms of the momentum space TMD FF and hemisphere soft function, respectively. 

We note that for this process, we have only considered a single-inclusive measurement in the hemisphere which contains the thrust axis, while the other plane is fully inclusive. For this type of measurement, only soft radiation which is emitted into the hemisphere containing the thrust axis will contribute to $j_\perp$. This subtlety introduces two complications which must be considered in the factorized expression. The first complication arises with the definition of the fully renormalized/finite so-called  properly  defined TMD FF. In the Collins-Soper-Sterman (CSS) treatment~\cite{Collins:2011zzd}
\begin{align}\label{e.TMDff}
    D^{\rm TMD}_{ \L/q}(z_{\L},b, \mu,\zeta)  = & D_{ \L/q}(z_{\L},b, \mu,\zeta/\nu^2) 
    \\ & \times \sqrt{S(b,\mu,\nu)}\,,\nn 
\end{align} 
where $S(b,\mu,\nu)$ is the standard soft function usually arose in the SIDIS, Drell-Yan and back-to-back hadron pair production in $e^+e^-$ collisions~\cite{Collins:2011zzd,Ji:2004wu,Chiu:2012ir,Echevarria:2015byo,Ebert:2019okf}. The explicit calculation of $S_{\rm hemi}$ given in \cite{Kang:2020yqw} demonstrated at one-loop order that
\bea
S_{\rm hemi}(b,\mu,\nu)  = \sqrt{S(b,\mu,\nu)}\,.
\eea
Because of this, the product of $D_{ \L/q}(z_{\L},b, \mu,\zeta/\nu^2)$ and $S_{\rm hemi}(b,\mu,\nu)$ in Eq.~\eqref{eq:unp-nll-FT} equals the standard TMD FF $D^{\rm TMD}_{ \L/q}(z_{\L},b, \mu,\zeta)$ in Eq.~\eqref{e.TMDff}. Thus the factorized expression in Eq.~\eqref{eq:unp-nll-FT} can be written as the following form
\begin{align}
    \label{eq:unp-nll-proper}
    \frac{d \sigma}{d z_{\L} d^2 \bm{j}_\perp} =\, \sigma_0 H(Q,\mu) & \sum_{q} e_q^2 \int_0^\infty \frac{d^2b}{(2\pi)^2} e^{i\bm{b}\cdot \bm{j}_\perp/z_\L}  \nn \\
    & \times D^{\rm TMD}_{ \L/q}(z_{\L},b, \mu, \zeta)  \,. 
\end{align}
We note that for all phenomenological applications, we will take $\mu^2 = \zeta = Q^2$ in the following discussions. Because of this, we suppress explicit $\mu$ and $\zeta$ dependence in our functions in this paper and instead give only explicit $Q$ dependence. 

The second complication that must be accounted for is that since we have restricted soft radiation to only one hemisphere, this observable is non-global \cite{Dasgupta:2001sh}. The factorization formula for non-global observables have been constructed in an effective field theory context in \cite{Becher:2015hka,Becher:2016mmh,Becher:2016omr,Becher:2017nof}, where a multi-Wilson-line structure \cite{Caron-Huot:2015bja,Nagy:2016pwq,Nagy:2017ggp} is the key ingredient to capture the non-linear QCD evolution effects from the so-called \textit{non-global logarithms}. It was recently shown in \cite{Kang:2020yqw}, that at NLL accuracy the factorization formula is given by
\begin{align}
\label{eq:unp-nll-NG1}
    \frac{d \sigma}{d z_{\L} d^2 \bm{j}_\perp} & =\, \sigma_0 \sum_{q} e_q^2 \int_0^\infty \frac{d^2b}{(2\pi)^2} e^{i\bm{b}\cdot \bm{j}_\perp/z_\L}  \\
    & \times D^{\rm TMD}_{ \L/q}(z_{\L},b, Q,Q^2) U_{\rm NG} (\mu_{b_*},Q)\nn \,. 
\end{align}
To arrive at this expression, we have introduced the auxiliary scale, $\mu_{b_*}$ which is defined in the $b_*$ prescription~\cite{Collins:1984kg}. We note that the most important difference between Eqs.~\eqref{eq:unp-nll-proper} and \eqref{eq:unp-nll-NG1} is the introduction of the function $U_{\rm NG}(\mu_{b_*},Q)$, which contains the effects of the non-global logarithms. Since the treatment of the non-global logarithms will be addressed phenomenologically in this paper, we will discuss this function in more detail in Sec.~\ref{TMD-Pheno}. 

In order to provide the full expression for the NLL cross section, the TMD FF must be matched onto the collinear fragmentation function through the relation
\begin{align}\label{eq:matching}
    D^{\rm TMD}_{ \L/q}(z_{\L},b, Q,Q^2) = & \frac{1}{z_{\L}^2} D_{ \L/q}(z_{\L},\mu_{b_*}) \\ 
    & \times e^{- S_{\rm pert}(\mu_{b_*},Q) -S_{\rm NP}(b,z_\L, Q_0, Q)}  \nn .
\end{align}
In order to arrive at this expression, we have performed tree level matching. The factor $S_{\rm NP}(b, z_\L,Q_0, Q)$ is the non-perturbative evolution factor for the unpolarized TMD FF where $Q_0$ is the initial TMD scale. Since this function depends on choice of parameterization, we will defer discussion of this function until \ref{TMD-Pheno}. On the other hand, the perturbative Sudakov factor is given by 
\begin{align}
S_{\rm pert}(\mu_b,Q) =& -\tilde K(b_*,\mu_{b_*})\ln\left(\frac{Q}{\mu_{b_*}}\right)\nonumber \\& -\int_{\mu_{b_{*}}}^{Q} \frac{d\mu}{\mu}\left[\gamma_F\left(\alpha_s(\mu), \frac{Q^2}{\mu^2}\right) \right]\,,
\end{align}
where at NLL order one has $K(b_*,\mu_{b_*})=0$ and 
\begin{align}
    & \gamma_F\left(\alpha_s(\mu), \frac{Q^2}{\mu^2}\right) = \frac{\alpha_s}{\pi}C_F  \left(\ln\frac{Q^2}{{\mu}^2} - \frac{3}{2}\right)\,
    \\
    & \hspace{1.5cm}+\frac{\alpha_s^2}{\pi^2} C_F \left[C_A\left(\frac{67}{18}-\frac{\pi^2}{6}\right)-\frac{10}{9} T_R\, n_f\right]\ln\frac{Q^2}{{\mu}^2}\nn\,.
\end{align}

To further simplify the expression for the differential cross section, it is now convenient to perform the integration over the $b$-space azimuthal angle. After performing this angular integration, we arrive at the final expression for the unpolarized scattering cross section at NLL
\begin{align}
\label{eq:unp-nll-final}
    \frac{d \sigma}{d z_{\L} d^2 \bm{j}_\perp} & =\, \sigma_0 \sum_{q} e_q^2 \int_0^\infty \frac{b db}{(2\pi)} J_0\left(\frac{b\, j_\perp}{z_\L} \right)  \\
    & \times \frac{1}{z_{\L}^2} D_{ \L/q}(z_{\L},\mu_{b_*}) \, e^{-S_{\rm NP}(b,z_\L, Q_0, Q) - S_{\rm pert}(\mu_{b_*},Q) } \nn \\ 
    & \times  U_{\rm NG} (\mu_{b_*},Q) \nn \,. 
\end{align}
In this expression, $J_0$ is the zero order Bessel function of the first kind. 

Now that we have summarized each of the pieces of the unpolarized cross section, we can extend this factorization theorem to the spin-dependent case. The spin-dependent differential cross section can be obtained from Eq.~\eqref{eq:unp-nll-NG1} by replacing the unpolarized TMD FF with the spin-dependent one. In order to obtain the $b$-space spin-dependent TMD FF, we begin by performing a Fourier transform of the momentum space spin-dependent TMD FF. In this paper, we follow the Trento conventions \cite{Bacchetta:2004jz} for the normalization of this function. This normalization is given by the expression,
\begin{align}
    \hat{D}_{ \L/q}& \big(z_{\L},\bm{p}_{\perp}, \bm{S}_\perp,Q \big) = \frac{1}{2}\Big{[} D_{\L/q}(z_{\L},p_{\L\perp},Q) \\
    & +\frac{1}{z_\L M_\L} D^{\perp}_{1T,\L/q}\left(z_{\L},p_{\perp},Q\right)\epsilon_{\perp \rho\sigma} p_{\perp}^{\rho} S_{\perp}^{\sigma} \Big{]}\nn\,.
\end{align}
The function $\hat{D}_{ \L/q} \big(z_{\L},\bm{p}_{\perp}, \bm{S}_\perp,Q \big)$ on the left hand side of this expression is the advertised spin-dependent TMD FF. The first term on the right hand side is the unpolarized TMD FF while the second term on the right hand side contains $D^{\perp}_{1T,\L/q}\left(z_{\L},p_{\L\perp},Q\right)$, the TMD PFF. We see in this expression that the first term on the right hand side is independent of the spin, while the second term in this expression is not. Therefore the first term in this expression contributes to the unpolarized cross section while the second contributes to the polarized cross section. 

After performing the Fourier transform of this expression, we arrive at the expression for the $b$-space spin-dependent TMD FF
\begin{align}
    \label{eq:bTMDFF}
    \hat{D}_{ \L/q}& \big(z_{\L},\bm{b}, \bm{S}_\perp,Q\big) = \frac{1}{2}\Big{[} D_{\L/q}(z_{\L},b,Q) \\
    & +\frac{iM_\L \epsilon_{\perp\rho\sigma} b^\rho S_{\perp}^{ \sigma}}{z_\L^2} D_{1T,\L/q}^{\perp(1)}\left(z_{\L},b,Q\right)\Big{]}\nn\,, 
\end{align}
In this expression, we have introduced the full spin-dependent $b$-space TMD FF
\begin{align}
\label{eq:FT}
    \hat{D}_{ \L/q} \big(z_\L,\bm{b}, \bm{S}_\perp,Q \big) = \frac{1}{z_\L^2}& \int d^2p_\perp e^{-i \bm{b}\cdot \bm{p}_\perp /z_\L} \\
    & \times \hat{D}_{ \L/q} \big(z_\L,\bm{p}_{\perp}, \bm{S}_\perp,Q \big)\nn\, ,
\end{align} 
as well as the $b$-space first Bessel moment-TMD PFF~\cite{Boer:2011xd}
\begin{align}
\label{eq:FT-D1T}
   & D_{1T,\L/q}^{\perp(1)} \left(z_{\L},b,Q\right)=-\frac{2z_\L^2}{M_\L^2}
   \frac{\partial }{\partial b^2} D_{1T,\L/q}^{\perp}\left(z_{\L},b,Q\right)\\ \nn
     &= \hspace{1mm}\frac{2\pi}{M_\L^2}\frac{z_\L^2}{b}
     \int \frac{dp_\perp}{z_\L}\left(\frac{p}{z}\right)^2 J_1\left(\frac{b\ p_\perp}{z_\L}\right) D_{1T,\L/q}^{\perp}\left(z_\L,p_{\perp},Q\right)\,. 
\end{align}
Analogous to the collinear matching of the TMD FF in Eq.~\eqref{eq:matching}, the TMD PFF can be matched to a collinear distribution, $D_{1T,\L/q}^{\perp(1)}\left(z_{\L},\mu_{b_*}\right)$ at NLL
\begin{align}
    \label{eq:matching-PFF}
 D_{1T,\L/q}^{\perp(1)}\left(z_{\L},b,Q\right)=  & 
 D_{1T,\L/q}^{\perp(1)}\left(z_{\L},\mu_{b_*}\right) \\ 
    & \times e^{ - S_{\rm pert}(\mu_{b_*},Q) -S_{\rm NP}^\perp(b,z_\L, Q_0, Q) } \nn \,,
\end{align}
which reduces to the ``transverse momentum'' moments~\cite{Mulders:1995dh,Boer:1997qn} in the small $b$ limit~\cite{Boer:2011xd}, 
\begin{align}
   \lim_{b\rightarrow 0}& D_{1T,\L/q}^{\perp(1)}\left(z_{\L},b,Q\right)  \\
    & = \frac{1}{M_\L^2}\frac{z_\L^2}{b}\int \frac{d^2p_\perp}{z_\L^2} \frac{p_\perp}{z_\L} \frac{bp_\perp}{2z_\L}D_{1T,\L/q}^{\perp}\left(z_\L,p_{\perp},Q\right) \nn \\
  & =\int d^2p_\perp  \frac{p_\perp^2}{z_\L^2 2M_\L^2}   D_{1T,\L/q}^{\perp}\left(z_\L,p_{\perp},Q\right) \nn\\
  &\equiv  D_{1T,\L/q}^{\perp(1)}\left(z_{\L},Q\right) \nn\, .
\end{align}

Furthermore, the non-perturbative evolution factor for the TMD PFF is denoted $S_{\rm NP}^\perp(b,z_\L, Q_0, Q)$. This non-perturbative factor is not the same as the unpolarized factor. We note that in order to make this difference clear, we have included a `$\perp$' in the superscript for the non-perturbative factor. The form of these functions will be addressed in \ref{TMD-Pheno}. Contrary to the non-perturbative evolution factor, the perturbative evolution factor, $S_{\rm pert}(\mu_{b_*},Q)$ is the same as the unpolarized case.

In order to arrive at an expression for the spin-dependent differential cross section, we now replace the unpolarized TMD FF in Eq.~\eqref{eq:unp-nll-NG1} with the spin-dependent TMD FF, 
\begin{align}
    \label{eq:all-nll}
    \frac{d \sigma\big(\bm{S}_\perp \big)}{d z_{\L} d^2 \bm{j}_\perp} & =\, \sigma_0 \sum_{q} e_q^2 \int_0^\infty \frac{d^2b}{(2\pi)^2} e^{i\bm{b}\cdot \bm{j}_\perp/z_\L}  \\
    & \hspace{0.5cm}\times \hat{D}_{ \L/q} \big(z_{\L},\bm{b}, \bm{S}_\perp,Q \big) U_{\rm NG} (\mu_{b_*},Q) \nn \,.
\end{align}
We can see from Eq.~\eqref{eq:bTMDFF} that the first term is independent of the transverse spin vector $S_\perp^\sigma$ while the second term is an odd function of $S_\perp^\sigma$. We can therefore isolate the unpolarized cross section by adding two full spin-dependent cross sections which have opposite spin configurations 
\begin{align}
    \frac{d \sigma}{d z_{\L} d^2 \bm{j}_\perp} & =\, \frac{d \sigma\big(\bm{S}_\perp \big)}{d z_{\L} d^2 \bm{j}_\perp}+\frac{d \sigma\big(\!-\!\bm{S}_\perp \big)}{d z_{\L} d^2 \bm{j}_\perp} \\
    \label{eq:TMD-upol}
    & =\, \sigma_0 \sum_{q} e_q^2 \int_0^\infty \frac{b db}{2\pi} J_0\left(\frac{b\, j_\perp}{z_\L} \right)  \\
    & \times \frac{1}{z_{\L}^2} D_{ \L/q}(z_{\L},\mu_{b_*}) \, e^{ -S_{\rm NP}(b,z_\L, Q_0, Q) - S_{\rm pert}(\mu_{b_*},Q) } \nn \\ 
    & \times  U_{\rm NG} (\mu_{b_*},Q) \nn \,. 
\end{align}
In order to isolate the contribution of the TMD PFF, we subtract two full spin-dependent cross sections which have opposite spin configurations.
\begin{align}
    & \frac{d\Delta \sigma}{d z_{\L} d^2 \bm{j}_\perp} = \frac{d \sigma\big(\bm{S}_\perp \big)}{d z_{\L} d^2 \bm{j}_\perp}-\frac{d \sigma\big(\!-\!\bm{S}_\perp \big)}{d z_{\L} d^2 \bm{j}_\perp}\label{eq:TMD-poll} \\
    & \hspace{10mm}=\, \sigma_0 \, \sin\left( \phi_s-\phi_j \right)\sum_{q} e_q^2 \int_0^\infty \frac{b^2db }{2\pi} J_1\left(\frac{bj_\perp}{z_\L}\right)  \nn \\
& \hspace{10mm} \times \frac{M_{\L}}{z^2_\L}
D_{1T,\L/q}^{\perp(1)}\left(z_{\L},\mu_{b_*}\right) e^{ -S_{\rm NP}^\perp(b,z_\L, Q_0, Q) - S_{\rm pert}(\mu_{b_*},Q) } \nn \\ 
    & \hspace{10mm} \times  U_{\rm NG} (\mu_{b_*},Q)  \, . \nn
\end{align}

To arrive at this expression, we have integrated over the $b$ azimuthal angle. From this expression, we see that the size of the spin-dependent cross section depends on the $\sin(\phi_s-\phi_j)$ modulation. In this modulation, the angles $\phi_s$ and $\phi_j$ are the azimuthal angles of $\bm S_\perp$ and $\bm j_\perp$, respectively. In Fig.~\ref{fig:thrust}, we provide a figure which demonstrates the definition of these angles relative to the other kinematics. In the experimentally measured polarization, it is conventional to take $\phi_s = \pi/2$ and $\phi_j = 0$ so that only the magnitude of the modulation is measured. For the purposes of this paper, we will always take these angles to be defined in this way. With this definition of the angles, the experimentally measured quantity for $\L$ transverse polarization is therefore given by the expression 
\begin{align}
    \label{eq:tmd-polarization}
    P_\perp^\L(z_\L,j_\perp) = \left.\frac{d\Delta \sigma}{d z_{\L} d^2 \bm{j}_\perp}\right/\frac{d \sigma}{d z_{\L} d^2 \bm{j}_\perp}\,.
\end{align}
\subsection{$\L$ Polarization in the CM Frame}\label{Twist3-Theory}
\begin{figure}[hbt!]
    \centering
    \includegraphics[width = 0.40\textwidth]{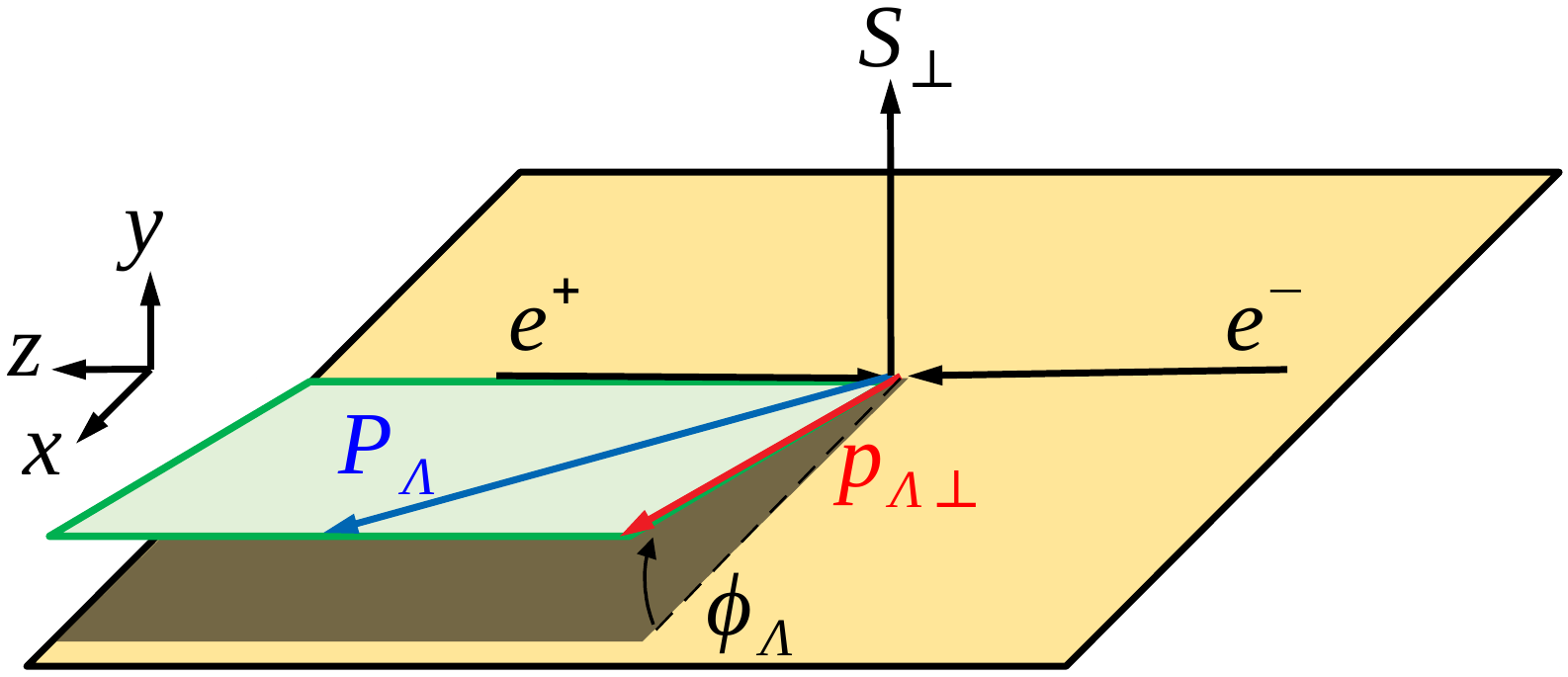}
    \caption{Transverse $\L$ polarization in the lepton CM frame.}
    \label{fig:twist}
\end{figure}
In this section, we consider the transverse polarization for the process
\begin{align}
    e^-(l) +e^+(l') \rightarrow \gamma^*(q) \rightarrow \L\big(z_\L, \bm{p}_{\L \perp}, \bm{S}_{\perp}\big) + X\,.
\end{align}
In this expression, $\bm {p}_{\L \perp}$ is the transverse momentum of the $\L$ baryon with respect to the lepton pair in the CM frame, while $z_\L$ and $S_\perp$ are defined in the same way as the previous section. For this process, each component of the fragmenting quark's momentum can be of order $Q$. Therefore each component of the $\L$ baryon's momentum, $P_\L$, is of order $z_\L Q$. Since each component of $P_\L$ is of the same order, a collinear factorization is well justified~\cite{Collins:1989gx,Bauer:2002nz,Collins:2011zzd,Gamberg:2018fwy}.

At LO, the collinear factorization for unpolarized single-inclusive hadron production has the well-known form
\begin{align}
\frac{E_\L d\sigma}{d^3 P_\L} = \frac{2 N_c \alpha_{\rm em}^2}{Q^4 z_\L}\left[ \left(1-v\right)^2+v^2 \right] \sum_{q} e_q^2  D_{\L/q}(z_\L,Q)\,.
\label{eq:Twist-3-dsigma}
\end{align}
In Eq.~\eqref{eq:Twist-3-dsigma}, we have introduced the kinematic variable $v = P_\L\cdot l'/P_\L\cdot q$ which is related to the $\L$ rapidity. 

For transversely polarized $\L$ production, the LO differential cross section was shown in \cite{Gamberg:2018fwy} to have the following form
\begin{align}
	\frac{E_\L d\Delta \sigma}{d^3 P_\L} = \frac{2 N_c \alpha_{\rm em}^2}{Q^4 z_\L}\frac{8 M_\L}{z_\L s^2}\epsilon^{\ell \ell' P_\L S_\perp}\frac{1}{z_\L}\sum_q e_q^2 D_{T,\L/q}(z_\L,Q)\,.
    \label{eq:twist-3Pol}
\end{align}
 In this expression
\begin{align}
    \epsilon^{\ell \ell' P_\L S_\perp} = \epsilon^{\mu \nu \rho \sigma} \ell_\mu \ell'_\nu P_{\L \rho} S_{\perp \sigma}\,.
\end{align} 
Here the four momentum of the $\L$ baryon is denoted $P_\L$ while the transverse spin of the $\L$ baryon is denoted $S_\perp$. In Fig.~\ref{fig:twist}, we provide a diagram which illustrates the kinematics for this process.  We note that a similar result to  Eq.~\eqref{eq:twist-3Pol} was given by Boer et al.~\cite{Boer:1997mf} in the context of a power suppressed one particle inclusive cross section  formalism.\footnote{It is of note that next to leading order factorization the collinear twist-3 formalism has been investigated in~\cite{Gamberg:2018fwy}, whereas no factorization for the twist-3 TMD framework has been proposed~\cite{Gamberg:2006ru,Bacchetta:2019qkv}.} We see that the relevant distribution function for this process is $D_{T,\L/q}(z_\L, Q)$, the intrinsic collinear twist-3 fragmentation function. 

In this paper, we will work in the lepton center of mass frame where $e^-$ and $e^+$ move in the positive and negative $z$ directions, respectively. In order to draw a clear connection to the TMD case, we choose to make the cross section differentiable in two kinematic parameters $\bm p_{\L \perp}$ and $z_\L$. After simplifying Eq.~\eqref{eq:Twist-3-dsigma}, the unpolarized cross section can be written as
\begin{align}
\label{eq:twist3-upol}
\frac{d\sigma}{dz_\L\, d^2p_{\L \perp}} & = \frac{2 N_c \alpha_{\rm em}^2}{Q^4 z_\L} \left(1-\frac{2 p_{\L \perp}^2}{z_\L ^2 Q^2}\right) \frac{Q}{2p_{\L z}} \nn \\
& \times \sum_{q} e_q^2 D_{\L/q}(z_\L,Q)\,
\end{align}
where 
\begin{align}
	p_{\L z} = \sqrt{\frac{Q^2}{4}z_\L^2-p_{\L \perp}^2} \,
\end{align}
is the magnitude of the $z$ component of $P_\L$. Since the magnitude of this component must be non-negative, we must have $p_{\L\perp} \leq Q z_\L/2$. Similarly, the transverse spin-dependent contribution to the cross section can be written as
\begin{align}
\label{eq:twist3-poll}
\frac{d\Delta \sigma}{dz_\L\, d^2 p_{\L \perp}}  = & -\sin{\left(\phi_s-\phi_\L\right)} \frac{2 N_c \alpha_{\rm em}^2}{Q^4 z_\L}  \left(\frac{4 M_\L}{Q}\right)\frac{p_{\L\perp}}{Q} \nn \\
& \times \frac{1}{z_\L^3} \sum_q e_q^2 \frac{D_{T,\L/q}(z_\L,Q)}{z_\L}\,.
\end{align}
Analogous to the TMD case, the transverse spin-dependent cross section is modulated by a factor of $\sin\left(\phi_s-\phi_\L\right)$. Here $\phi_s$ and $\phi_\L$ are the azimuthal angles of the spin vector, $\bm S_\perp$, and the $\L$ baryon transverse momentum, $\bm p_{\L \perp}$, respectively. In Fig.~\ref{fig:twist}, these angles are shown with respect to the other kinematic variables in the measurement. Experimental measurements of the asymmetry will usually take the convention that $\phi_s = \pi/2$ and $\phi_\L = 0$. In our paper, we will always follow this convention. After setting the values of these angles, the polarization in the twist-3 formalism is given by
\begin{align}\label{eq:pLCM}
    P_{\rm \textbf{CM}}^\L(z_\L,p_{\L \perp}) = \left.\frac{d\Delta \sigma}{dz_\L\, d^2 p_{\L \perp}}\right/\frac{d\sigma}{dz_\L\, d^2 p_{\L \perp}}\,.
\end{align}
At this point, it is important to note that the polarization in the CM frame is proportional to $M_\L/Q$. 

\section{Phenomenology}\label{Pheno}
In this section, we first use the TMD formalism in the previous section to compute the polarization in the thrust frame and compare with the OPAL and Belle measurements. We then make a prediction for polarization for the polarization in the CM frame at Belle kinematics. 
\subsection{TMD Phenomenology}\label{TMD-Pheno}
As we saw in Sec.~\ref{TMD-Theory}, the denominator of the TMD polarization at NLL is given in Eq.~\eqref{eq:TMD-upol}. In this paper, we will use the standard $b_*$ prescription from \cite{Collins:1984kg}
\begin{align}
b_* = \frac{b}{\sqrt{1+b^2/b_{\rm max}^2}}\,,
\end{align}
where $b_{\rm max}$ characterizes the boundary between the non-perturbative and perturbative regions for $b$ dependence~\cite{Collins:2011qcdbook,Collins:2014jpa}. Typical values used in phenomenology range from approximate $0.5 \lesssim b_{\rm max} \lesssim 1.5$~\cite{Konychev:2005iy,Collins:2015dfa}. In order to describe the collinear FF, we follow the work in \cite{Callos:2020qtu} to use the collinear AKK fragmentation function \cite{Albino:2008fy} for $D_{\L/q}\left(z_\L,\mu_{b_{*}} \right)$. Note that the AKK fragmentation function in \cite{Albino:2008fy} is only given in the region with $Q > Q_{\rm min} = 1$ GeV. Thus, since $\mu_{b*} > Q_{\rm min}$~\cite{Konychev:2005iy} this restricts  $b_{\rm max} \lesssim  1.1$ GeV$^{-1}$ using the AKK fragmentation functions; we choose $b_{\rm max}=0.5$.

Furthermore, for the non-perturbative function $S_{\rm NP}(b,z_\L, Q_0, Q)$ 
we use the parametrization~\cite{Aidala:2014hva,Su:2014wpa}
\begin{align}
    \label{eq:parameter}
    S_{\rm NP}(b,z_\L, Q_0,Q) = g_h \frac{b^2}{z_\L^2}+\frac{g_2}{2}\ln{\frac{Q}{Q_0}}\ln{\frac{b}{b_*}}  \,
\end{align}
for the fragmentation function. Here there are two non-perturbative parameters, $g_h$ and $g_2$. The parameter $g_h$ controls the Gaussian width of the unpolarized TMD FF at the initial scale $Q_0$, with $g_h \simeq \langle p_\perp^2\rangle/4$. On the other hand, the parameter $g_2$ is universal for all TMDs~\cite{Collins:2011qcdbook,Collins:2014jpa} and controls the evolution from $Q_0$ to $Q$. In order to obtain numerical values for $g_h$ and $Q_0$, we closely follow the parameterization in \cite{Callos:2020qtu}, where the Gaussian width $\langle p_\perp^2\rangle$ is translated to $g_h \simeq 0.048$ GeV$^2$ at $Q_0 = 10.58$ GeV. Furthermore, we use the value of $g_2 = 0.84$ which was obtained in \cite{Su:2014wpa} from a global fit from unpolarized SIDIS and Drell-Yan data.

In order to account for the non-linear QCD evolution associated with the non-global logarithms, we follow the parameterization in \cite{Dasgupta:2001sh} 
\begin{align}
U_{\rm NG}(\mu_{b_*},Q) = \exp\left[ -C_A C_F \frac{\pi^2}{3}u^2 \frac{1+(au)^2}{1+(bu)^c} \right]
\end{align}
with $a = 0.85 C_A$, $b = 0.86 C_A$, $c = 1.33$ and
\begin{align}
	u \equiv \int_{\mu_{b_*}}^{Q}\frac{d\mu}{\mu}\frac{\alpha_s(\mu)}{2\pi} = \frac{1}{\beta_0}\ln\left[\frac{\alpha_s(\mu_{b_*})}{\alpha_s(Q)}\right]
\end{align}
$\beta_0 = \frac{11}{3}C_A-\frac{4}{3}T_F n_f$, with $T_F = 1/2$. Finally, in order to perform the numerical Bessel transform in Eq.~\eqref{eq:TMD-upol}, we use the numerical algorithm in \cite{Kang:2019ctl}. 

Having summarized the details for the unpolarized scattering cross section, we will provide the details for the transverse spin-dependent cross section. In Sec.~\ref{TMD-Theory}, the polarized differential cross section was shown to be given in Eq.~\eqref{eq:TMD-poll}. In order to obtain a parameterization for the $b$-space TMD PFF, we note that the momentum space TMD PFF was recently extracted at the scale $Q_0 = 10.58$ GeV in \cite{Callos:2020qtu} using the parameterization
\begin{align}
\label{eq:TMDPFF-gauss}
D_{1T,h/q}^\perp(z,p_\perp,Q_0) = 
  D_{1T,h/q}^{\perp}(z,Q_0)\frac{e^{-p_{\perp}^2/\langle M_D^2\rangle}}{\pi \langle M_D^2\rangle}\,.
\end{align}
In this expressions, $D_{1T,h/q}^{\perp}(z,Q_0)$ is the collinear PFF~\cite{Callos:2020qtu} while $\langle M_D^2 \rangle$ is the Gaussian width which was extracted from the Belle data. In this reference, the authors take the parameterization 
\begin{align}
D_{1T,h/q}^{\perp}(z,Q_0) = \mathcal{N}_q(z)D_{h/q}(z,Q_0)\,.
\end{align}
Here the factor $\mathcal{N}_q(z)$ is a collinear modulation function which is parameterized by the expression
\begin{align}
\mathcal{N}_q(z) = N_q z^{\alpha_q}(1-z)^{\beta_q}\frac{(\alpha_q+\beta_q-1)^{\alpha_q+\beta_q-1}}{(\alpha_q-1)^{\alpha_q-1}\beta_q^{\beta_q}}\,.
\end{align}
The parameters $\alpha_q$, $\beta_q$, and $N_q$ were all determined from the fit in~\cite{Callos:2020qtu}.

Using this parameterization, we obtain the following expression for the $b$-space TMD PFF in Eq.~\eqref{eq:matching-PFF} with QCD evolution
\begin{align}
    D_{1T,\L/q}^{\perp(1)}\left(z,b,Q\right) 
    &=\frac{\langle M_D^2\rangle}{2z^2 M_\L^2}
D_{1T,\L/q}^{\perp}  (z,\mu_{b_*})
   \nonumber\\
    &\times e^{ - S_{\rm pert}(\mu_{b_*},Q)-S_{\rm NP}^\perp(b,z,Q_0, Q)}
\end{align}
where
\begin{align}
S_{\rm NP}^\perp\left(b,z,Q_0,Q\right) = \frac{\langle M_D^2\rangle}{4} \frac{b^2}{z^2}+\frac{g_2}{2}\ln{\frac{Q}{Q_0}}\ln{\frac{b}{b_*}}\,.
\end{align}
We used again that the coefficient in front of $b^2$ is given by the corresponding Gaussian width, $\langle M_D^2\rangle/4$, and $g_2$ is universal for all TMDs. 
Now that we have supplied all of the details of the phenomenology, we can compare our theoretical prediction in Eq.~\eqref{eq:tmd-polarization} against the OPAL and Belle data. 

\begin{figure}[hbt!]
    \centering
    \includegraphics[width = 0.3\textwidth]{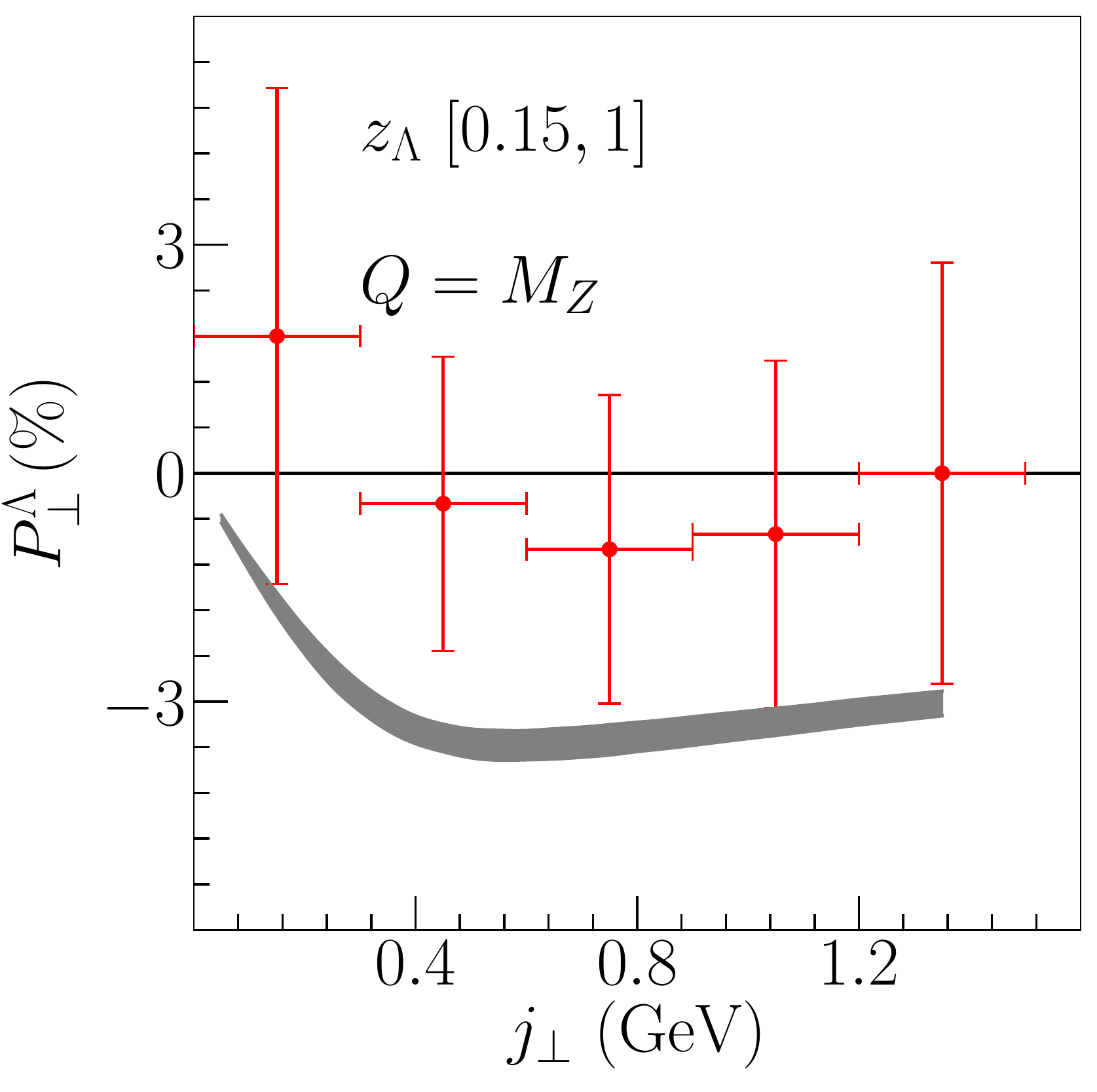}
    \caption{$P_\perp^\L(z_\L,j_\perp)$ in $e^+\,e^-\rightarrow \L(\rm Thrust)\,X$ for OPAL \cite{Ackerstaff:1997nh}. The theoretical curve is integrated over the region $0.15<z_\L<1$. We plot the experimental data in red with the total experimental uncertainty as a vertical error bar while the experimental uncertainty on $j_\perp$ is in the horizontal error bar. The gray band is the theoretical uncertainty which was generated from the replicas for the TMD PFF in~\cite{Callos:2020qtu}.}
    \label{TMD-LEP}
\end{figure}
\begin{figure*}[hbt!]
    \centering
    \includegraphics[width = \textwidth]{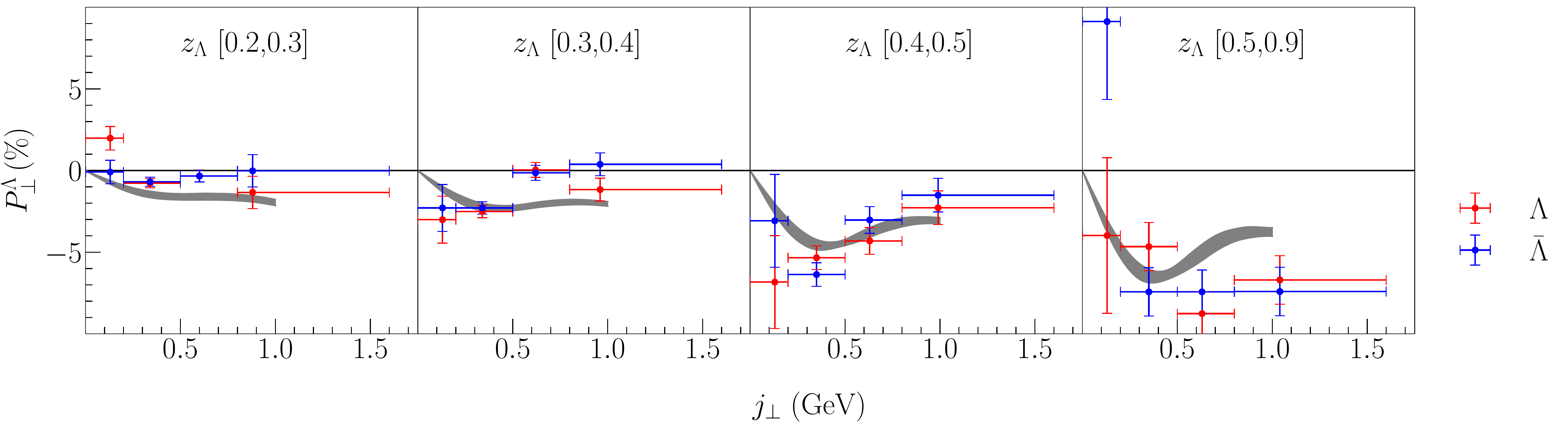}
    \caption{$P_\perp^\L(z_\L,j_\perp)$ in $e^+\,e^-\rightarrow \L(\rm Thrust)\,X$ for the Belle data \cite{Guan:2018ckx}. From left column to right column, the theoretical curve is integrated from $0.2<z_\L<0.3$, $0.3<z_\L<0.4$, $0.4<z_\L<0.5$, $0.5<z_\L<0.9$. The data in red is for $\L$ production while the data in blue is for $\bar{\L}$ production. The experimental data is plotted with the total experimental uncertainty as a vertical error bar while the experimental uncertainty on $j_\perp$ is in the horizontal error bar. The gray band represents the theoretical prediction with the uncertainty obtained from the replicas for the TMD PFF.}
    \label{TMD}
\end{figure*}

In Fig.~\ref{TMD-LEP}, we plot the polarization as a function of $j_\perp$. We note that our convention for the direction of the vector $S_\perp$ is opposite of the direction that was used in the OPAL measurement. To account for this different convention, we have multiplied the experimental data by a minus sign. We also note that the experimental data at OPAL is integrated over the region $0.15 \leq z_\L \leq 1$. In our calculation, we have also included the theoretical uncertainty from the fit performed in \cite{Callos:2020qtu}. To generate this theoretical uncertainty, we have generated a theoretical prediction for each of the 201 replicas in \cite{Callos:2020qtu}. At each data point, we have a set of 201 predictions and we keep the middle $68\%$ of this set by cutting the bottom and top $16$ percentile. The band is then generated from the maximum to the minimum of this cut set. This uncertainty is plotted as a gray band in our description of the experimental data.
While our theoretical description of the data is slightly larger than the central values of the experimental  result,  we expect that the OPAL data can be used in a future global analysis to constrain the form and evolution of the TMD PFF.

In Fig.~\ref{TMD}, we plot our theoretical calculations against the Belle data. The columns from left to right of this figure indicated the binned values for the $z_\L$ that we used in our numerical calculations. To generate our theoretical curve, we integrate over the advertised $z_\L$ values. It is important to note that the rightmost bin in the experimental data was $0.5 \leq z_\L \leq 0.9$. While the TMD PFF in \cite{Callos:2020qtu} was extracted in the region $0.2\leq z_\L \leq 0.5$, we also provide our prediction for the final bin. In this plot, the blue data is for $\L$ production while the red data is for $\bar{\L}$ production. The horizontal error bars indicate the bin size in $j_\perp$ while the vertical error bars are the total experimental error. We note that the TMD PFF in our phenomenology is invariant under charge conjugation, explicitly $D^\perp_{1T,\L/q}(z,b,Q) = D^\perp_{1T,\bar{\L}/\bar{q}}(z,b,Q)$. Therefore, after performing the sum over the quark flavors, the theoretical prediction for $\L$ and $\bar{\L}$ is then the same. We see in Fig.~\ref{TMD} that in the region of small $z_\L$, the magnitude of the experimental data is small. This behavior can be described by examining Fig.~5 in \cite{Callos:2020qtu}. At small $z_\L$ the magnitude of the $u$, $d$, and sea TMD PFFs are large and the sign of the $u$ TMD PFF is opposite of the $d$ and sea TMD PFFs. Therefore in this region there are large cancellations that are occurring between the different flavors. However, at $z_\L>0.4$, the $d$ and $s$ TMD PFFs dominate. Since the $d$ and $s$ quark TMD PFFs have the same sign, the magnitude of the theoretical curve is larger in that region. We see in the regions $0.2\leq z_\L \leq 0.3$, $0.3\leq z_\L \leq 0.4$, and $0.4\leq z_\L \leq 0.5$ that our theoretical prediction agrees with the experimental data. Furthermore, while the TMD PFF was only extracted in the region $0.2\leq z_\L \leq 0.5$, we find that the parameterization still describes the experimental data well in the region $0.5<z_\L <0.9$.

\subsection{Twist-3 Phenomenology}\label{Twist-Pheno}
In this section, we provide our prediction for the twist-3 transverse polarization at Belle. The denominator for the twist-3 polarization is given by Eq.~\eqref{eq:twist3-upol}. In order to generate a numerical prediction for unpolarized $\L$ production, we only need to fix the collinear unpolarized FFs for $\L$ baryons. For this purpose, we once again use the AKK collinear FFs in~\cite{Albino:2008fy}. 

\begin{figure}
    \centering    
    \includegraphics[height = 0.33\textwidth]{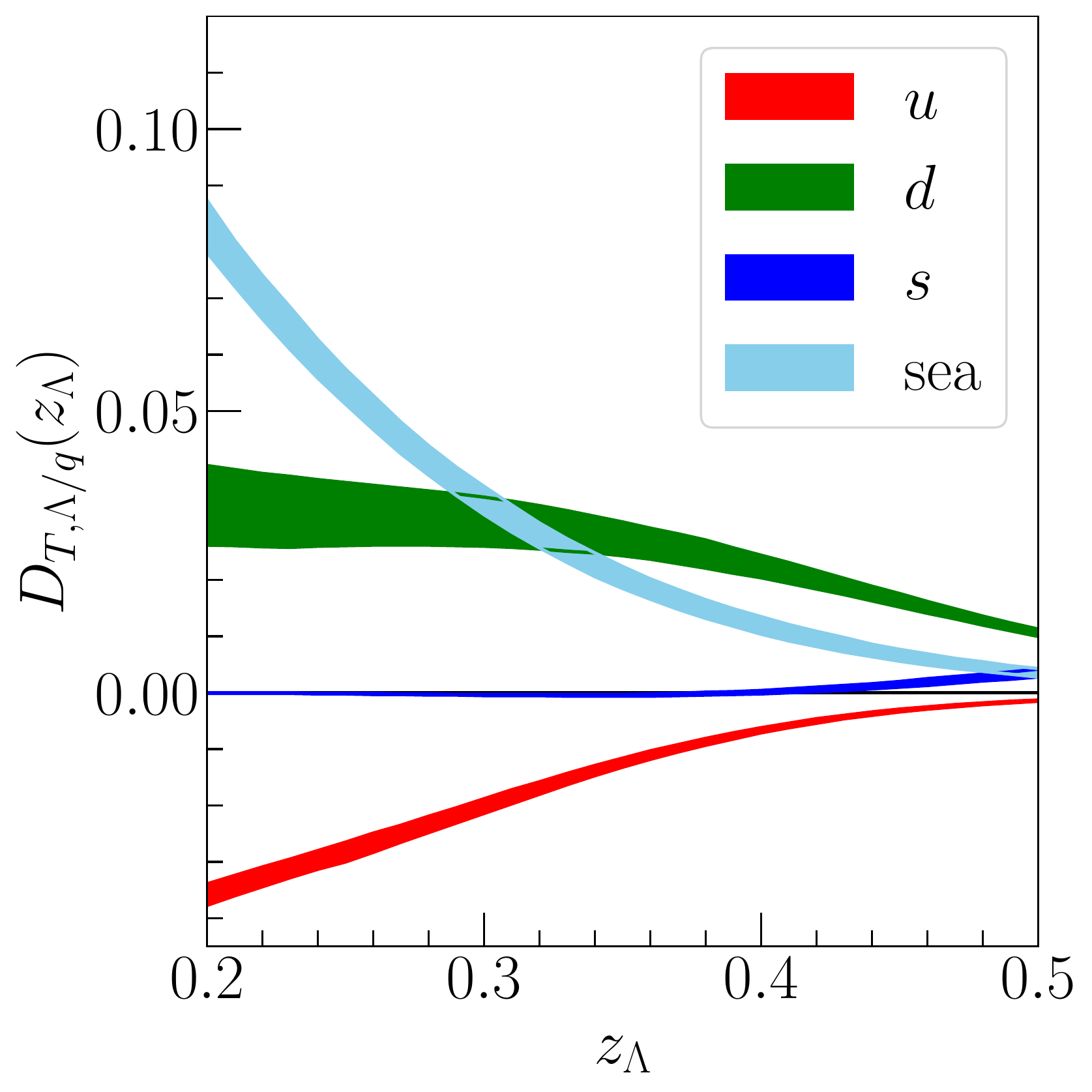}
    \caption{The twist-3 fragmentation functions $D_{T,\L /q}(z_\L, Q)$, defined in Eq.~\eqref{eqn:dtWW}, plotted as functions of $z_\L$. The bands are the $68\%$ confidence region while the line is the central curve.}
    \label{fig:dt}
\end{figure}

On the other hand, we saw that the numerator of the polarization is given in Eq.~\eqref{eq:twist3-poll}. Therefore in order to describe this process, we only need a parameterization for $D_{T,\L/q}(z_\L,Q)$. Given our lack of knowledge of this fundamental twist-3 T-odd fragmentation function, we will employ the approach outlined in~\cite{Gamberg:2017jha} in order to re-express the $D_{T,\L/q}(z_\L, Q)$ in terms of our knowledge of $D_{1T,\L/q}^{\perp (1)}(z_\L, Q)$. We observe that we can relate  this intrinsic twist-3 FF  to the kinematic and dynamical twist-3 functions~\cite{Metz:2012ct} through the relation,  
\begin{align}
\frac{1}{z_\L}D_{T,\L/q}(z_\L, Q) = & -\left(1-z_\L\frac{d}{dz_\L}\right)
D_{1T,\L/q}^{\perp(1)}(z_\L, Q) 
	\nn \\
	& -2 \int_0^1 d\beta \frac{\Im\left[ \hat{D}_{FT}^{qg} (z_\L, Q, \beta)\right]}{(1-\beta)^2} \,.
    \label{eqn:dt}
\end{align}
which as derived in Ref.~\cite{Kanazawa:2015ajw} by employing both 
Lorentz invariance relations and equations of motion relations (EOMs). In this expression $D_{1T,\L/q}^{\perp (1)}(z_\L, Q)$ is the kinematic twist-3 fragmentation function which is defined in terms of the TMD PFF in the previous section through the relation
\begin{align}\label{eq:kinematic3}
D_{1T,\L/q}^{\perp (1)}(z_\L,Q) & = \int d^2 p_{\L\perp} \frac{p_{\L\perp}^2}{2z_\L^2 M_\L^2} D_{1T,\L/q}^\perp\left(z,p_{\L \perp}^2,Q \right)\, ,
\end{align}
where a regularization procedure is implied~\cite{Collins:2016hqq,Gamberg:2017jha,Qiu:2020oqr}.
That is, the collinear limit of the first Bessel moment of the TMD PFF in Eq.~\eqref{eq:matching-PFF} corresponds to  the first moment of the TMD PFF, through a  limiting procedure, as $b$ becomes very small and is associated with the hard scale $b\sim 1/Q$~\cite{Collins:2016hqq,Gamberg:2017jha}, resulting in a renormalized first moment of the functions originally introduced in~\cite{Mulders:1995dh,Boer:1997mf}. 

On the other hand, $\hat{D}_{FT}^{qg}(z_\L,Q, \beta)$ is the dynamical twist-3 fragmentation function~\cite{Metz:2012ct}. 
Since from Eq.~\eqref{eq:kinematic3}  the kinematic function $D_{1T,\L/q}^{\perp (1)}(z_\L,Q)$ is related to the TMD PFFs,  we can use the extracted results from \cite{Callos:2020qtu} to obtain this function. The dynamical twist-3 function $\hat{D}_{FT}^{qg}(z,Q,\beta)$ on the other hand is not yet known. 
In order to perform a phenomenological analysis, we adopt the approach  outlined in Ref.~\cite{Gamberg:2017jha} where as a first approximation we neglect the last term in Eq.~\eqref{eqn:dt}; that is,
\begin{align}\label{eqn:dtWW}
	\frac{1}{z_\L}D_{T,\L/q}(z_\L,Q) = & -\left( 1-z_\L \frac{d}{dz_\L}\right)D_{1T,\L/q}^{\perp (1)}(z_\L,Q)\,.
\end{align}
This is a statement that integral in Eq.~\eqref{eqn:dt} is parametrically smaller than the first term: not that $\hat{D}_{FT}^{qg}(z_\L,\beta)$  is zero. 
In fact $\hat{D}_{FT}^{qg}(z_\L,Q, \beta)$ must not be zero since it was shown in Ref.~\cite{Kanazawa:2015ajw},  
$D_{1T,\L/q}^{\perp (1)}(z_\L,Q)$ is an integral of $\hat{D}_{FT}^{qg}(z_\L,Q, \beta)$. Our purpose here is to 
provide  first estimate of  
$D_{T,\L/q}(z_\L, Q)$ in order to provide a first prediction of  
$P_{\rm \textbf{CM}}^\L(z_\L,p_{\L \perp})$ in $e^+\,e^-\rightarrow \L\,X$ for the Belle data~\cite{Guan:2018ckx}. Once data from Belle is analyzed in the CM frame, we can ascertain the size of the neglected contribution.

In Fig.~\ref{fig:dt} we plot the function $D_{T,\L/q}(z_\L, Q)$ computed in Eq.~\eqref{eqn:dtWW} as a function of $z_\L$ at $Q = 10.58$ GeV for $u$, $d$, $s$, and the sea quarks. To generate the contribution of the sea quarks, we add the contributions of the $\bar{u}$, $\bar{d}$, $\bar{s}$-quarks. We once again follow the procedure which is given in detail in the previous section to generate the uncertainty band. In red we plot $D_{T,\L/u}(z_\L, Q)$ while in green we plot $D_{T,\L/d}(z_\L, Q)$. In blue, we plot  $D_{T,\L/s}(z_\L, Q)$. Finally in light blue, we plot the sum over the sea quarks. We find that at small $z_\L$, the sum over the sea quarks for $D_{T,\L/q}(z_\L, Q)$ becomes very large. However at large $z_\L$, the contribution of the sea quarks become small. At small $z_\L$, the next largest contribution comes from the $u$-quarks which are negative. On the other hand, the contribution of the $d$-quarks are positive. At small $z_\L$ the contribution from the $d$ quarks is smaller than that of the sea and $u$ quarks. However, at large $z_\L$, the $d$-quark has the largest contribution. We find that $D_{T,\L/s}(z_\L, Q)$ tends to be much smaller than the other twist-3 FFs in the plotted region.

\begin{figure}[t]
    \includegraphics[height = 0.22\textwidth]{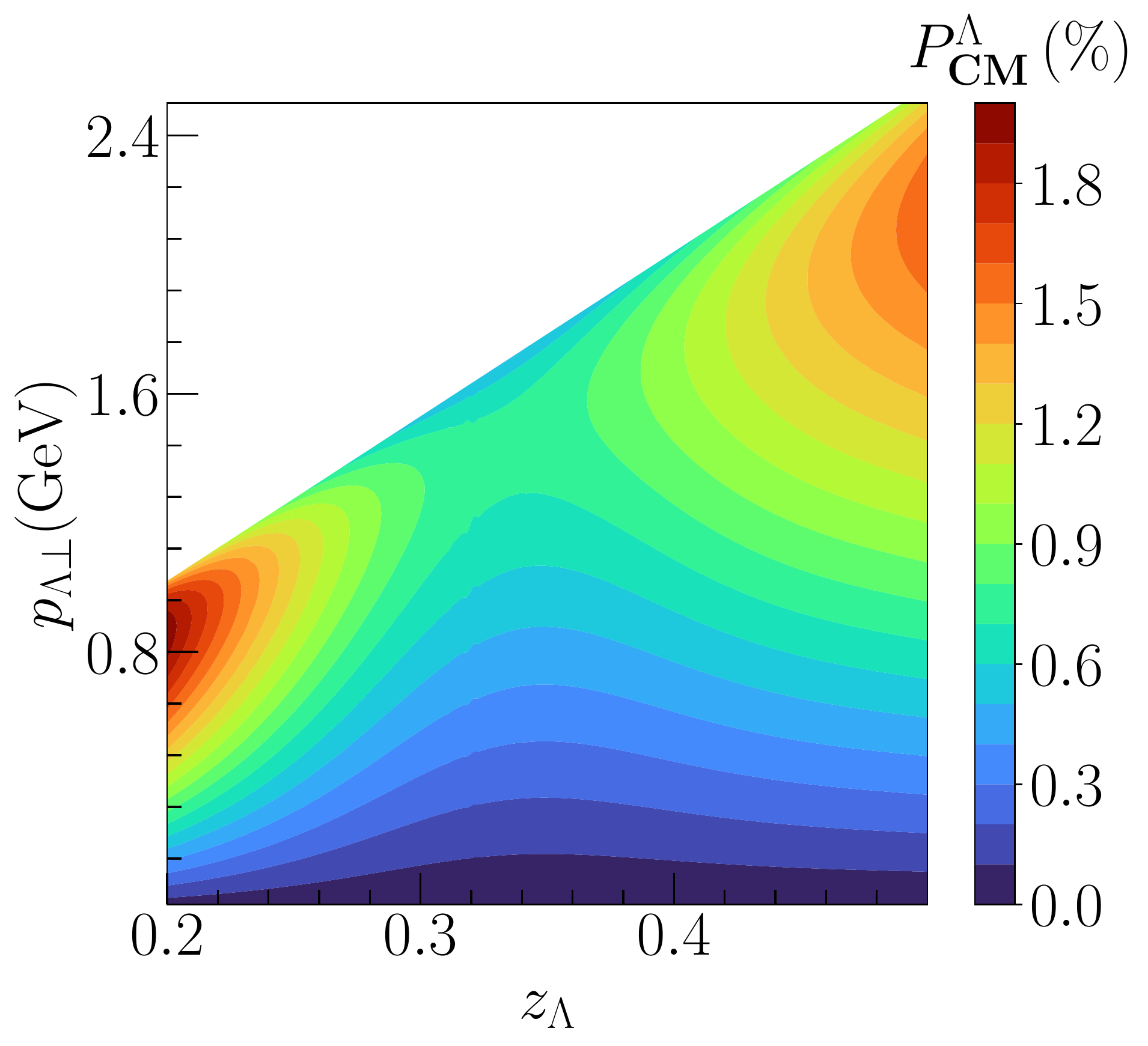}
    \includegraphics[height = 0.22\textwidth]{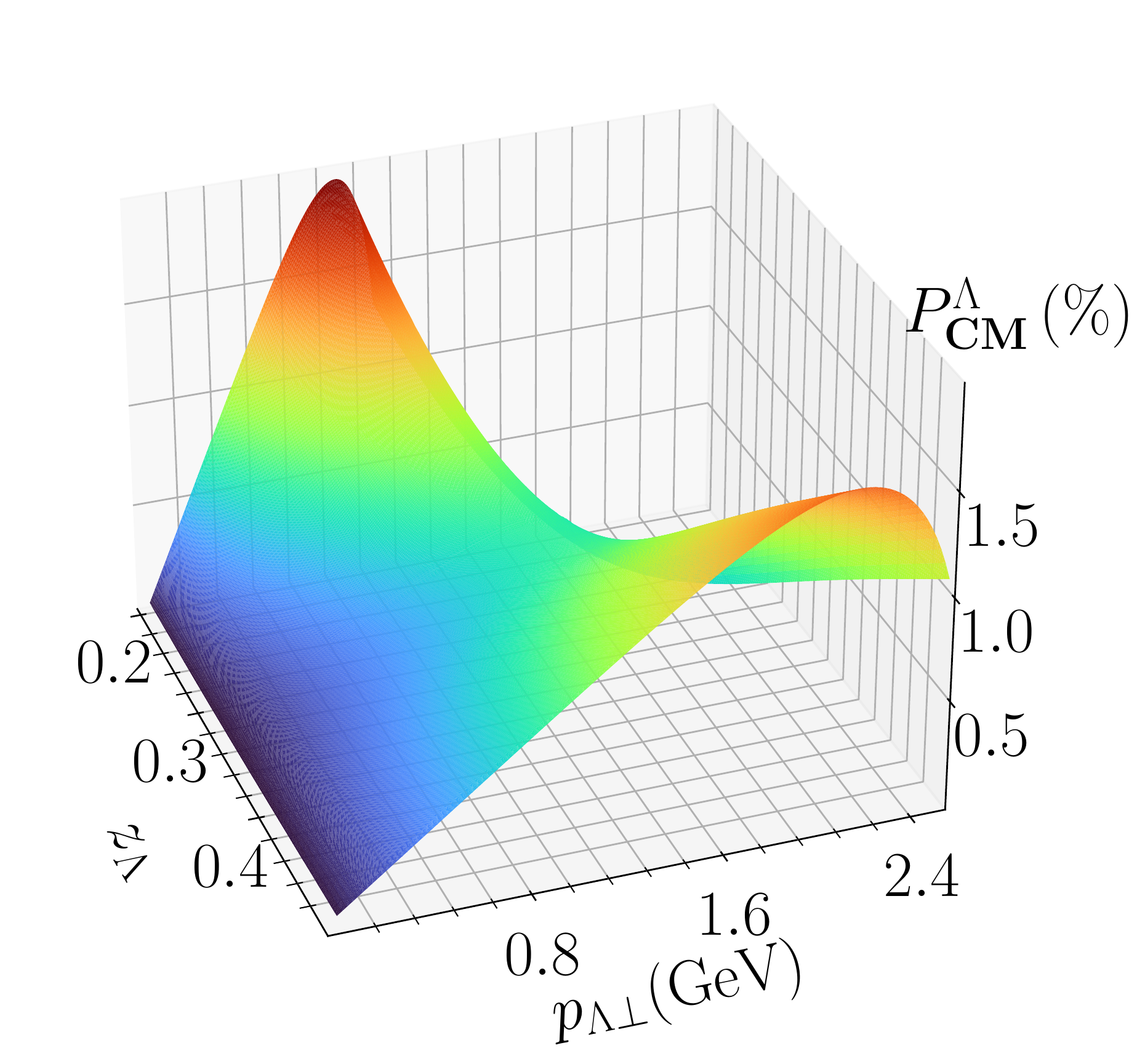}
    \includegraphics[width = 0.47\textwidth]{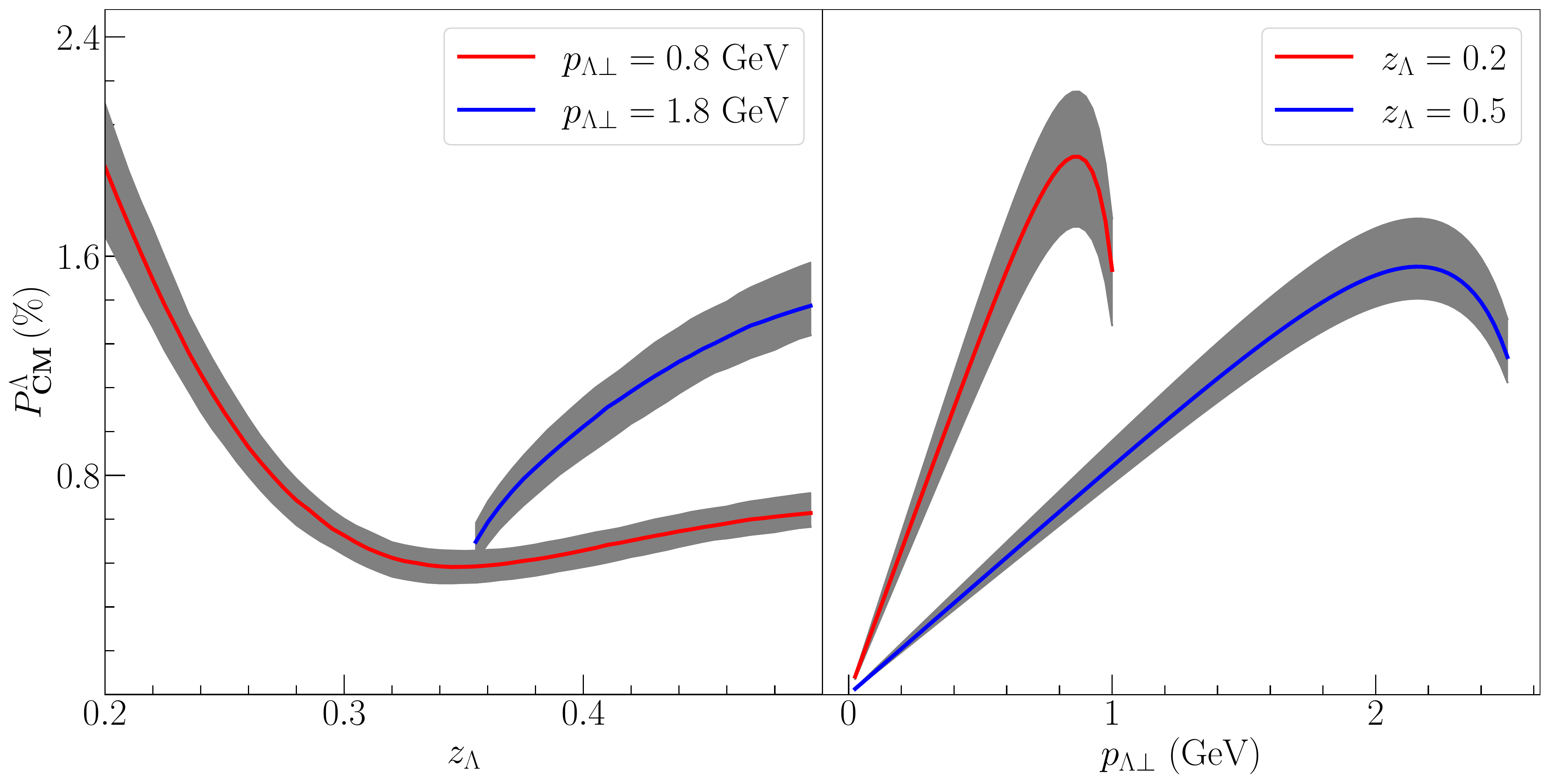}
    \caption{$P_{\rm \textbf{CM}}^\L(z_\L,p_{\L \perp})$ defined in Eq.~\eqref{eq:pLCM} for $e^+e^-\to \L X$ as a function of $z_\L$ and $p_{\L\perp}$ at Belle. Top left: A contour plot of the polarization in $z_\L$ and $p_{\L \perp}$. Top right: A three dimensional plot of the polarization. Bottom left: Plot of the polarization as a function of $z_\L$ only. Bottom right: Plot of the polarization as a function of $p_{\L \perp}$ only. The top plots are generated only using the central fit. The red and blue curves in the bottom plots are generated using the central fit while the gray band is the theoretical uncertainty.}
    \label{fig:three graphs}
\end{figure}

In order to generate a prediction for Belle~\cite{Guan:2018ckx}, in Fig.~\ref{fig:three graphs} we plot the polarization as a function of both $p_{\L\perp}$ and $z_\L$ at $Q = 10.58$ GeV. We note that because the $p_{\L z}$ must be non-negative, we then only plot our prediction in the region where $p_{\L\perp}\leq z_\L Q/2$. Therefore as $z_\L$ grows, the range of available $p_{\L \perp}$ increases. Furthermore due to this phase space restriction, at $Q = 10.58$ GeV the transverse momentum of the $\L$ baryon can then be at most a few GeV. However, it is important to note that this transverse momentum originates from the hard interaction as well as the collinear fragmentation, and not from any TMD physics. In the top-left plot of this figure, we provide a contour plot for $P_{\rm \textbf{CM}}^\L(z_\L,p_{\L \perp})$ in Eq.~\eqref{eq:pLCM} in $e^+e^-\to \L X$ using the central fit parameter values. In the top-right, we provide a three-dimensional plot of the polarization using the central fit. In the bottom-left plot, we show the polarization as a function of $z_\L$ at $p_{\L\perp} = 0.8$ GeV in blue and $p_{\L\perp} = 1.8$ GeV in red. In the bottom-right plot, we see the polarization as a function of $p_{\L \perp}$ for $z_\L = 0.2$ and $z_\L = 0.5$. In all of our plots, we see that the size of the predicted asymmetry tends to be $1-2\%$ in magnitude. While the size of this polarization is relatively small, it is important to note that this prediction was made using a Wandzura-Wilczek type relation, Eq.~\eqref{eqn:dtWW}. If the size of the polarization at Belle is found to be larger than our prediction, this could be an indication that the Wandzura-Wilczek relation is not a good approximation  for this function. Furthermore, if a significant signal for this process is found, this would also be the first demonstration that the function $D_{T, \L/q}$ is non-zero. 

Additionally, we have made theoretical predictions against the measurement at OPAL. As shown in \cite{Ackerstaff:1997nh}, the polarization $P_{\rm \textbf{CM}}^\L(z_\L,p_{\L \perp})$ measured in the direction of $\hat{S}_\perp=\hat{z}\times\hat{P}_\L$ is given to be $1.1\pm1.8\%$ for $z_\L>0.15$ and $p_{\L\perp}>0.3$ GeV/$c$. With the theoretical framework provided in this work, we obtain $P_{\rm \textbf{CM}}^\L=0.020^{+0.003}_{-0.004}\%$ with $z_\L$ integrated from $0.15$ to $1$ and $p_{\L\perp}$ integrated from $0.3$ GeV to $z_\L Q/2$, which is in agreement with the measurement provided by OPAL.

\section{Conclusion}\label{Conclusion}

In this paper we have studied transverse polarization in single-inclusive $\L$ production in a thrust TMD framework as well as a collinear twist-3 framework. We have shown that when the polarization is differential in the transverse momentum with respect to the thrust axis, the polarization can be used to probe the TMD Polarizing Fragmentation Function (PFF), $D_{1T}^\perp$. On the other hand, we have also shown that when polarization is differential in the transverse momentum with respect to the leptons, the polarization is sensitive to the $D_{T,\L}$ function, a collinear twist-3 fragmentation function. In order to describe the first process, we have extended a recent TMD formalism. Using this formalism, we have generated a theoretical prediction and compared this against the OPAL and Belle measurements. We have found good theoretical description of the Belle thrust axis data, and also demonstrated that the
OPAL is reasonably described from the thrust axis factorization theorem developed here. This analysis provides proof of principle that the experimental data at OPAL and Belle can be described using the factorization and resummation formalism that we have introduced. Future work could involve using these experimental data to constrain the evolution of the TMD PFF.

Furthermore we have discussed how this recent Belle data can in principle be re-binned to measure the transverse momentum of the $\L$ baryon with respect to the lepton pair. Using a collinear twist-3 formalism, we have generated a theoretical prediction for this polarization at Belle. This measurement will allow for the first measurement of the intrinsic twist-3 FF, $D_{T,\L/q}$ which by naive time reversal~\cite{DeRujula:1972te, Collins:1992kk,Goeke:2005hb,Metz:2016swz} symmetry, is predicted to be non-zero.

\section*{Acknowledgements}
L.G. is supported by the US Department of Energy under
contract No.~DE-FG02-07ER41460. Z.K., D.Y.S. and F.Z. are supported by the National Science Foundation under Grant No.~PHY-1945471. J.T. is supported by NSF Graduate  Research Fellowship Program under Grant No.~DGE-1650604. D.Y.S. is also supported by Center for Frontiers in Nuclear Science of Stony Brook University and Brookhaven National Laboratory. This work is supported within the framework of the TMD Topical Collaboration.

\bibliography{refs}

\begin{thebibliography}{78}%
\makeatletter
\providecommand \@ifxundefined [1]{%
 \@ifx{#1\undefined}
}%
\providecommand \@ifnum [1]{%
 \ifnum #1\expandafter \@firstoftwo
 \else \expandafter \@secondoftwo
 \fi
}%
\providecommand \@ifx [1]{%
 \ifx #1\expandafter \@firstoftwo
 \else \expandafter \@secondoftwo
 \fi
}%
\providecommand \natexlab [1]{#1}%
\providecommand \enquote  [1]{``#1''}%
\providecommand \bibnamefont  [1]{#1}%
\providecommand \bibfnamefont [1]{#1}%
\providecommand \citenamefont [1]{#1}%
\providecommand \href@noop [0]{\@secondoftwo}%
\providecommand \href [0]{\begingroup \@sanitize@url \@href}%
\providecommand \@href[1]{\@@startlink{#1}\@@href}%
\providecommand \@@href[1]{\endgroup#1\@@endlink}%
\providecommand \@sanitize@url [0]{\catcode `\\12\catcode `\$12\catcode
  `\&12\catcode `\#12\catcode `\^12\catcode `\_12\catcode `\%12\relax}%
\providecommand \@@startlink[1]{}%
\providecommand \@@endlink[0]{}%
\providecommand \url  [0]{\begingroup\@sanitize@url \@url }%
\providecommand \@url [1]{\endgroup\@href {#1}{\urlprefix }}%
\providecommand \urlprefix  [0]{URL }%
\providecommand \Eprint [0]{\href }%
\providecommand \doibase [0]{http://dx.doi.org/}%
\providecommand \selectlanguage [0]{\@gobble}%
\providecommand \bibinfo  [0]{\@secondoftwo}%
\providecommand \bibfield  [0]{\@secondoftwo}%
\providecommand \translation [1]{[#1]}%
\providecommand \BibitemOpen [0]{}%
\providecommand \bibitemStop [0]{}%
\providecommand \bibitemNoStop [0]{.\EOS\space}%
\providecommand \EOS [0]{\spacefactor3000\relax}%
\providecommand \BibitemShut  [1]{\csname bibitem#1\endcsname}%
\let\auto@bib@innerbib\@empty
\bibitem [{\citenamefont {Bunce}\ \emph {et~al.}(1976)\citenamefont {Bunce}
  \emph {et~al.}}]{Bunce:1976yb}%
  \BibitemOpen
  \bibfield  {author} {\bibinfo {author} {\bibfnamefont {G.}~\bibnamefont
  {Bunce}} \emph {et~al.},\ }\href {\doibase 10.1103/PhysRevLett.36.1113}
  {\bibfield  {journal} {\bibinfo  {journal} {Phys. Rev. Lett.}\ }\textbf
  {\bibinfo {volume} {36}},\ \bibinfo {pages} {1113} (\bibinfo {year}
  {1976})}\BibitemShut {NoStop}%
\bibitem [{\citenamefont {Schachinger}\ \emph {et~al.}(1978)\citenamefont
  {Schachinger} \emph {et~al.}}]{Schachinger:1978qs}%
  \BibitemOpen
  \bibfield  {author} {\bibinfo {author} {\bibfnamefont {L.}~\bibnamefont
  {Schachinger}} \emph {et~al.},\ }\href {\doibase 10.1103/PhysRevLett.41.1348}
  {\bibfield  {journal} {\bibinfo  {journal} {Phys. Rev. Lett.}\ }\textbf
  {\bibinfo {volume} {41}},\ \bibinfo {pages} {1348} (\bibinfo {year}
  {1978})}\BibitemShut {NoStop}%
\bibitem [{\citenamefont {Heller}\ \emph {et~al.}(1983)\citenamefont {Heller}
  \emph {et~al.}}]{Heller:1983ia}%
  \BibitemOpen
  \bibfield  {author} {\bibinfo {author} {\bibfnamefont {K.~J.}\ \bibnamefont
  {Heller}} \emph {et~al.},\ }\href {\doibase 10.1103/PhysRevLett.51.2025}
  {\bibfield  {journal} {\bibinfo  {journal} {Phys. Rev. Lett.}\ }\textbf
  {\bibinfo {volume} {51}},\ \bibinfo {pages} {2025} (\bibinfo {year}
  {1983})}\BibitemShut {NoStop}%
\bibitem [{\citenamefont {Kane}\ \emph {et~al.}(1978)\citenamefont {Kane},
  \citenamefont {Pumplin},\ and\ \citenamefont {Repko}}]{Kane:1978nd}%
  \BibitemOpen
  \bibfield  {author} {\bibinfo {author} {\bibfnamefont {G.~L.}\ \bibnamefont
  {Kane}}, \bibinfo {author} {\bibfnamefont {J.}~\bibnamefont {Pumplin}}, \
  and\ \bibinfo {author} {\bibfnamefont {W.}~\bibnamefont {Repko}},\ }\href
  {\doibase 10.1103/PhysRevLett.41.1689} {\bibfield  {journal} {\bibinfo
  {journal} {Phys. Rev. Lett.}\ }\textbf {\bibinfo {volume} {41}},\ \bibinfo
  {pages} {1689} (\bibinfo {year} {1978})}\BibitemShut {NoStop}%
\bibitem [{\citenamefont {Lundberg}\ \emph {et~al.}(1989)\citenamefont
  {Lundberg} \emph {et~al.}}]{Lundberg:1989hw}%
  \BibitemOpen
  \bibfield  {author} {\bibinfo {author} {\bibfnamefont {B.}~\bibnamefont
  {Lundberg}} \emph {et~al.},\ }\href {\doibase 10.1103/PhysRevD.40.3557}
  {\bibfield  {journal} {\bibinfo  {journal} {Phys. Rev.}\ }\textbf {\bibinfo
  {volume} {D40}},\ \bibinfo {pages} {3557} (\bibinfo {year}
  {1989})}\BibitemShut {NoStop}%
\bibitem [{\citenamefont {Yuldashev}\ \emph {et~al.}(1991)\citenamefont
  {Yuldashev} \emph {et~al.}}]{Yuldashev:1990az}%
  \BibitemOpen
  \bibfield  {author} {\bibinfo {author} {\bibfnamefont {B.~S.}\ \bibnamefont
  {Yuldashev}} \emph {et~al.},\ }\href {\doibase 10.1103/PhysRevD.43.2792}
  {\bibfield  {journal} {\bibinfo  {journal} {Phys. Rev.}\ }\textbf {\bibinfo
  {volume} {D43}},\ \bibinfo {pages} {2792} (\bibinfo {year}
  {1991})}\BibitemShut {NoStop}%
\bibitem [{\citenamefont {Ramberg}\ \emph {et~al.}(1994)\citenamefont {Ramberg}
  \emph {et~al.}}]{Ramberg:1994tk}%
  \BibitemOpen
  \bibfield  {author} {\bibinfo {author} {\bibfnamefont {E.~J.}\ \bibnamefont
  {Ramberg}} \emph {et~al.},\ }\href {\doibase 10.1016/0370-2693(94)91397-8}
  {\bibfield  {journal} {\bibinfo  {journal} {Phys. Lett.}\ }\textbf {\bibinfo
  {volume} {B338}},\ \bibinfo {pages} {403} (\bibinfo {year}
  {1994})}\BibitemShut {NoStop}%
\bibitem [{\citenamefont {Panagiotou}(1990)}]{Panagiotou:1989sv}%
  \BibitemOpen
  \bibfield  {author} {\bibinfo {author} {\bibfnamefont {A.~D.}\ \bibnamefont
  {Panagiotou}},\ }\href {\doibase 10.1142/S0217751X90000568} {\bibfield
  {journal} {\bibinfo  {journal} {Int. J. Mod. Phys. A}\ }\textbf {\bibinfo
  {volume} {5}},\ \bibinfo {pages} {1197} (\bibinfo {year} {1990})}\BibitemShut
  {NoStop}%
\bibitem [{\citenamefont {Dharmaratna}\ and\ \citenamefont
  {Goldstein}(1996)}]{Dharmaratna:1996xd}%
  \BibitemOpen
  \bibfield  {author} {\bibinfo {author} {\bibfnamefont {W.~G.}\ \bibnamefont
  {Dharmaratna}}\ and\ \bibinfo {author} {\bibfnamefont {G.~R.}\ \bibnamefont
  {Goldstein}},\ }\href {\doibase 10.1103/PhysRevD.53.1073} {\bibfield
  {journal} {\bibinfo  {journal} {Phys. Rev. D}\ }\textbf {\bibinfo {volume}
  {53}},\ \bibinfo {pages} {1073} (\bibinfo {year} {1996})}\BibitemShut
  {NoStop}%
\bibitem [{\citenamefont {Anselmino}\ \emph {et~al.}(2001)\citenamefont
  {Anselmino}, \citenamefont {Boer}, \citenamefont {Dalesio},\ and\
  \citenamefont {Murgia}}]{Anselmino:2001la}%
  \BibitemOpen
  \bibfield  {author} {\bibinfo {author} {\bibfnamefont {M.}~\bibnamefont
  {Anselmino}}, \bibinfo {author} {\bibfnamefont {D.}~\bibnamefont {Boer}},
  \bibinfo {author} {\bibfnamefont {U.}~\bibnamefont {Dalesio}}, \ and\
  \bibinfo {author} {\bibfnamefont {F.}~\bibnamefont {Murgia}},\ }\href@noop {}
  {\bibfield  {journal} {\bibinfo  {journal} {Phys. Rev.}\ }\textbf {\bibinfo
  {volume} {D63}},\ \bibinfo {pages} {054029} (\bibinfo {year}
  {2001})}\BibitemShut {NoStop}%
\bibitem [{\citenamefont {Anselmino}\ \emph {et~al.}(2002)\citenamefont
  {Anselmino}, \citenamefont {Boer}, \citenamefont {D'Alesio},\ and\
  \citenamefont {Murgia}}]{Anselmino:2001js}%
  \BibitemOpen
  \bibfield  {author} {\bibinfo {author} {\bibfnamefont {M.}~\bibnamefont
  {Anselmino}}, \bibinfo {author} {\bibfnamefont {D.}~\bibnamefont {Boer}},
  \bibinfo {author} {\bibfnamefont {U.}~\bibnamefont {D'Alesio}}, \ and\
  \bibinfo {author} {\bibfnamefont {F.}~\bibnamefont {Murgia}},\ }\href
  {\doibase 10.1103/PhysRevD.65.114014} {\bibfield  {journal} {\bibinfo
  {journal} {Phys. Rev.}\ }\textbf {\bibinfo {volume} {D65}},\ \bibinfo {pages}
  {114014} (\bibinfo {year} {2002})},\ \Eprint
  {http://arxiv.org/abs/hep-ph/0109186} {arXiv:hep-ph/0109186 [hep-ph]}
  \BibitemShut {NoStop}%
\bibitem [{\citenamefont {Boer}\ \emph {et~al.}(2010)\citenamefont {Boer},
  \citenamefont {Kang}, \citenamefont {Vogelsang},\ and\ \citenamefont
  {Yuan}}]{Boer:2010ya}%
  \BibitemOpen
  \bibfield  {author} {\bibinfo {author} {\bibfnamefont {D.}~\bibnamefont
  {Boer}}, \bibinfo {author} {\bibfnamefont {Z.-B.}\ \bibnamefont {Kang}},
  \bibinfo {author} {\bibfnamefont {W.}~\bibnamefont {Vogelsang}}, \ and\
  \bibinfo {author} {\bibfnamefont {F.}~\bibnamefont {Yuan}},\ }\href {\doibase
  10.1103/PhysRevLett.105.202001} {\bibfield  {journal} {\bibinfo  {journal}
  {Phys. Rev. Lett.}\ }\textbf {\bibinfo {volume} {105}},\ \bibinfo {pages}
  {202001} (\bibinfo {year} {2010})},\ \Eprint {http://arxiv.org/abs/1008.3543}
  {arXiv:1008.3543 [hep-ph]} \BibitemShut {NoStop}%
\bibitem [{\citenamefont {Boer}(2010)}]{Boer:2010yp}%
  \BibitemOpen
  \bibfield  {author} {\bibinfo {author} {\bibfnamefont {D.}~\bibnamefont
  {Boer}},\ }\href {\doibase 10.22323/1.106.0215} {\bibfield  {journal}
  {\bibinfo  {journal} {PoS}\ }\textbf {\bibinfo {volume} {DIS2010}},\ \bibinfo
  {pages} {215} (\bibinfo {year} {2010})},\ \Eprint
  {http://arxiv.org/abs/1007.3145} {arXiv:1007.3145 [hep-ph]} \BibitemShut
  {NoStop}%
\bibitem [{\citenamefont {Wei}\ \emph {et~al.}(2015)\citenamefont {Wei},
  \citenamefont {Chen}, \citenamefont {Song},\ and\ \citenamefont
  {Liang}}]{Wei:2014pma}%
  \BibitemOpen
  \bibfield  {author} {\bibinfo {author} {\bibfnamefont {S.-Y.}\ \bibnamefont
  {Wei}}, \bibinfo {author} {\bibfnamefont {K.-b.}\ \bibnamefont {Chen}},
  \bibinfo {author} {\bibfnamefont {Y.-k.}\ \bibnamefont {Song}}, \ and\
  \bibinfo {author} {\bibfnamefont {Z.-t.}\ \bibnamefont {Liang}},\ }\href
  {\doibase 10.1103/PhysRevD.91.034015} {\bibfield  {journal} {\bibinfo
  {journal} {Phys. Rev. D}\ }\textbf {\bibinfo {volume} {91}},\ \bibinfo
  {pages} {034015} (\bibinfo {year} {2015})},\ \Eprint
  {http://arxiv.org/abs/1410.4314} {arXiv:1410.4314 [hep-ph]} \BibitemShut
  {NoStop}%
\bibitem [{\citenamefont {Gamberg}\ \emph {et~al.}(2019)\citenamefont
  {Gamberg}, \citenamefont {Kang}, \citenamefont {Pitonyak}, \citenamefont
  {Schlegel},\ and\ \citenamefont {Yoshida}}]{Gamberg:2018fwy}%
  \BibitemOpen
  \bibfield  {author} {\bibinfo {author} {\bibfnamefont {L.}~\bibnamefont
  {Gamberg}}, \bibinfo {author} {\bibfnamefont {Z.-B.}\ \bibnamefont {Kang}},
  \bibinfo {author} {\bibfnamefont {D.}~\bibnamefont {Pitonyak}}, \bibinfo
  {author} {\bibfnamefont {M.}~\bibnamefont {Schlegel}}, \ and\ \bibinfo
  {author} {\bibfnamefont {S.}~\bibnamefont {Yoshida}},\ }\href {\doibase
  10.1007/JHEP01(2019)111} {\bibfield  {journal} {\bibinfo  {journal} {JHEP}\
  }\textbf {\bibinfo {volume} {01}},\ \bibinfo {pages} {111} (\bibinfo {year}
  {2019})},\ \Eprint {http://arxiv.org/abs/1810.08645} {arXiv:1810.08645
  [hep-ph]} \BibitemShut {NoStop}%
\bibitem [{\citenamefont {Fanti}\ \emph {et~al.}(1999)\citenamefont {Fanti}
  \emph {et~al.}}]{Fanti:1998px}%
  \BibitemOpen
  \bibfield  {author} {\bibinfo {author} {\bibfnamefont {V.}~\bibnamefont
  {Fanti}} \emph {et~al.},\ }\href {\doibase 10.1007/s100520050337} {\bibfield
  {journal} {\bibinfo  {journal} {Eur. Phys. J.}\ }\textbf {\bibinfo {volume}
  {C6}},\ \bibinfo {pages} {265} (\bibinfo {year} {1999})}\BibitemShut
  {NoStop}%
\bibitem [{\citenamefont {Abt}\ \emph {et~al.}(2006)\citenamefont {Abt} \emph
  {et~al.}}]{Abt:2006da}%
  \BibitemOpen
  \bibfield  {author} {\bibinfo {author} {\bibfnamefont {I.}~\bibnamefont
  {Abt}} \emph {et~al.} (\bibinfo {collaboration} {HERA-B}),\ }\href {\doibase
  10.1016/j.physletb.2006.05.040} {\bibfield  {journal} {\bibinfo  {journal}
  {Phys. Lett.}\ }\textbf {\bibinfo {volume} {B638}},\ \bibinfo {pages} {415}
  (\bibinfo {year} {2006})},\ \Eprint {http://arxiv.org/abs/hep-ex/0603047}
  {arXiv:hep-ex/0603047 [hep-ex]} \BibitemShut {NoStop}%
\bibitem [{\citenamefont {Erhan}\ \emph {et~al.}(1979)\citenamefont {Erhan}
  \emph {et~al.}}]{Erhan:1979xm}%
  \BibitemOpen
  \bibfield  {author} {\bibinfo {author} {\bibfnamefont {S.}~\bibnamefont
  {Erhan}} \emph {et~al.},\ }\href {\doibase 10.1016/0370-2693(79)90761-5}
  {\bibfield  {journal} {\bibinfo  {journal} {Phys. Lett.}\ }\textbf {\bibinfo
  {volume} {82B}},\ \bibinfo {pages} {301} (\bibinfo {year}
  {1979})}\BibitemShut {NoStop}%
\bibitem [{\citenamefont {Aad}\ \emph {et~al.}(2015)\citenamefont {Aad} \emph
  {et~al.}}]{ATLAS:2014ona}%
  \BibitemOpen
  \bibfield  {author} {\bibinfo {author} {\bibfnamefont {G.}~\bibnamefont
  {Aad}} \emph {et~al.} (\bibinfo {collaboration} {ATLAS}),\ }\href {\doibase
  10.1103/PhysRevD.91.032004} {\bibfield  {journal} {\bibinfo  {journal} {Phys.
  Rev.}\ }\textbf {\bibinfo {volume} {D91}},\ \bibinfo {pages} {032004}
  (\bibinfo {year} {2015})},\ \Eprint {http://arxiv.org/abs/1412.1692}
  {arXiv:1412.1692 [hep-ex]} \BibitemShut {NoStop}%
\bibitem [{\citenamefont {Ackerstaff}\ \emph {et~al.}(1998)\citenamefont
  {Ackerstaff} \emph {et~al.}}]{Ackerstaff:1997nh}%
  \BibitemOpen
  \bibfield  {author} {\bibinfo {author} {\bibfnamefont {K.}~\bibnamefont
  {Ackerstaff}} \emph {et~al.} (\bibinfo {collaboration} {OPAL}),\ }\href
  {\doibase 10.1007/s100520050123} {\bibfield  {journal} {\bibinfo  {journal}
  {Eur. Phys. J. C}\ }\textbf {\bibinfo {volume} {2}},\ \bibinfo {pages} {49}
  (\bibinfo {year} {1998})},\ \Eprint {http://arxiv.org/abs/hep-ex/9708027}
  {arXiv:hep-ex/9708027} \BibitemShut {NoStop}%
\bibitem [{\citenamefont {Guan}\ \emph {et~al.}(2019)\citenamefont {Guan} \emph
  {et~al.}}]{Guan:2018ckx}%
  \BibitemOpen
  \bibfield  {author} {\bibinfo {author} {\bibfnamefont {Y.}~\bibnamefont
  {Guan}} \emph {et~al.} (\bibinfo {collaboration} {Belle}),\ }\href {\doibase
  10.1103/PhysRevLett.122.042001} {\bibfield  {journal} {\bibinfo  {journal}
  {Phys. Rev. Lett.}\ }\textbf {\bibinfo {volume} {122}},\ \bibinfo {pages}
  {042001} (\bibinfo {year} {2019})},\ \Eprint
  {http://arxiv.org/abs/1808.05000} {arXiv:1808.05000 [hep-ex]} \BibitemShut
  {NoStop}%
\bibitem [{\citenamefont {Mulders}\ and\ \citenamefont
  {Tangerman}(1996)}]{Mulders:1995dh}%
  \BibitemOpen
  \bibfield  {author} {\bibinfo {author} {\bibfnamefont {P.}~\bibnamefont
  {Mulders}}\ and\ \bibinfo {author} {\bibfnamefont {R.}~\bibnamefont
  {Tangerman}},\ }\href {\doibase 10.1016/0550-3213(95)00632-X} {\bibfield
  {journal} {\bibinfo  {journal} {Nucl. Phys. B}\ }\textbf {\bibinfo {volume}
  {461}},\ \bibinfo {pages} {197} (\bibinfo {year} {1996})},\ \bibinfo {note}
  {[Erratum: Nucl.Phys.B 484, 538--540 (1997)]},\ \Eprint
  {http://arxiv.org/abs/hep-ph/9510301} {arXiv:hep-ph/9510301} \BibitemShut
  {NoStop}%
\bibitem [{\citenamefont {Collins}\ and\ \citenamefont
  {Soper}(1981)}]{Collins:1981uk}%
  \BibitemOpen
  \bibfield  {author} {\bibinfo {author} {\bibfnamefont {J.~C.}\ \bibnamefont
  {Collins}}\ and\ \bibinfo {author} {\bibfnamefont {D.~E.}\ \bibnamefont
  {Soper}},\ }\href {\doibase 10.1016/0550-3213(81)90339-4} {\bibfield
  {journal} {\bibinfo  {journal} {Nucl. Phys. B}\ }\textbf {\bibinfo {volume}
  {193}},\ \bibinfo {pages} {381} (\bibinfo {year} {1981})},\ \bibinfo {note}
  {[Erratum: Nucl.Phys.B 213, 545 (1983)]}\BibitemShut {NoStop}%
\bibitem [{\citenamefont {Boer}\ \emph {et~al.}(1997)\citenamefont {Boer},
  \citenamefont {Jakob},\ and\ \citenamefont {Mulders}}]{Boer:1997mf}%
  \BibitemOpen
  \bibfield  {author} {\bibinfo {author} {\bibfnamefont {D.}~\bibnamefont
  {Boer}}, \bibinfo {author} {\bibfnamefont {R.}~\bibnamefont {Jakob}}, \ and\
  \bibinfo {author} {\bibfnamefont {P.}~\bibnamefont {Mulders}},\ }\href
  {\doibase 10.1016/S0550-3213(97)00456-2} {\bibfield  {journal} {\bibinfo
  {journal} {Nucl. Phys. B}\ }\textbf {\bibinfo {volume} {504}},\ \bibinfo
  {pages} {345} (\bibinfo {year} {1997})},\ \Eprint
  {http://arxiv.org/abs/hep-ph/9702281} {arXiv:hep-ph/9702281} \BibitemShut
  {NoStop}%
\bibitem [{\citenamefont {Collins}(2013)}]{Collins:2011zzd}%
  \BibitemOpen
  \bibfield  {author} {\bibinfo {author} {\bibfnamefont {J.}~\bibnamefont
  {Collins}},\ }\href@noop {} {\emph {\bibinfo {title} {{Foundations of
  perturbative QCD}}}},\ Vol.~\bibinfo {volume} {32}\ (\bibinfo  {publisher}
  {Cambridge University Press},\ \bibinfo {year} {2013})\BibitemShut {NoStop}%
\bibitem [{\citenamefont {Collins}(1993)}]{Collins:1992kk}%
  \BibitemOpen
  \bibfield  {author} {\bibinfo {author} {\bibfnamefont {J.~C.}\ \bibnamefont
  {Collins}},\ }\href {\doibase 10.1016/0550-3213(93)90262-N} {\bibfield
  {journal} {\bibinfo  {journal} {Nucl. Phys. B}\ }\textbf {\bibinfo {volume}
  {396}},\ \bibinfo {pages} {161} (\bibinfo {year} {1993})},\ \Eprint
  {http://arxiv.org/abs/hep-ph/9208213} {arXiv:hep-ph/9208213} \BibitemShut
  {NoStop}%
\bibitem [{\citenamefont {Metz}(2002)}]{Metz:2002iz}%
  \BibitemOpen
  \bibfield  {author} {\bibinfo {author} {\bibfnamefont {A.}~\bibnamefont
  {Metz}},\ }\href {\doibase 10.1016/S0370-2693(02)02899-X} {\bibfield
  {journal} {\bibinfo  {journal} {Phys. Lett. B}\ }\textbf {\bibinfo {volume}
  {549}},\ \bibinfo {pages} {139} (\bibinfo {year} {2002})},\ \Eprint
  {http://arxiv.org/abs/hep-ph/0209054} {arXiv:hep-ph/0209054} \BibitemShut
  {NoStop}%
\bibitem [{\citenamefont {Collins}\ and\ \citenamefont
  {Metz}(2004)}]{Collins:2004nx}%
  \BibitemOpen
  \bibfield  {author} {\bibinfo {author} {\bibfnamefont {J.~C.}\ \bibnamefont
  {Collins}}\ and\ \bibinfo {author} {\bibfnamefont {A.}~\bibnamefont {Metz}},\
  }\href {\doibase 10.1103/PhysRevLett.93.252001} {\bibfield  {journal}
  {\bibinfo  {journal} {Phys. Rev. Lett.}\ }\textbf {\bibinfo {volume} {93}},\
  \bibinfo {pages} {252001} (\bibinfo {year} {2004})},\ \Eprint
  {http://arxiv.org/abs/hep-ph/0408249} {arXiv:hep-ph/0408249} \BibitemShut
  {NoStop}%
\bibitem [{\citenamefont {Meissner}\ and\ \citenamefont
  {Metz}(2009)}]{Meissner:2008yf}%
  \BibitemOpen
  \bibfield  {author} {\bibinfo {author} {\bibfnamefont {S.}~\bibnamefont
  {Meissner}}\ and\ \bibinfo {author} {\bibfnamefont {A.}~\bibnamefont
  {Metz}},\ }\href {\doibase 10.1103/PhysRevLett.102.172003} {\bibfield
  {journal} {\bibinfo  {journal} {Phys. Rev. Lett.}\ }\textbf {\bibinfo
  {volume} {102}},\ \bibinfo {pages} {172003} (\bibinfo {year} {2009})},\
  \Eprint {http://arxiv.org/abs/0812.3783} {arXiv:0812.3783 [hep-ph]}
  \BibitemShut {NoStop}%
\bibitem [{\citenamefont {Gamberg}\ \emph {et~al.}(2011)\citenamefont
  {Gamberg}, \citenamefont {Mukherjee},\ and\ \citenamefont
  {Mulders}}]{Gamberg:2010uw}%
  \BibitemOpen
  \bibfield  {author} {\bibinfo {author} {\bibfnamefont {L.~P.}\ \bibnamefont
  {Gamberg}}, \bibinfo {author} {\bibfnamefont {A.}~\bibnamefont {Mukherjee}},
  \ and\ \bibinfo {author} {\bibfnamefont {P.~J.}\ \bibnamefont {Mulders}},\
  }\href {\doibase 10.1103/PhysRevD.83.071503} {\bibfield  {journal} {\bibinfo
  {journal} {Phys. Rev. D}\ }\textbf {\bibinfo {volume} {83}},\ \bibinfo
  {pages} {071503} (\bibinfo {year} {2011})},\ \Eprint
  {http://arxiv.org/abs/1010.4556} {arXiv:1010.4556 [hep-ph]} \BibitemShut
  {NoStop}%
\bibitem [{\citenamefont {D'Alesio}\ \emph {et~al.}(2020)\citenamefont
  {D'Alesio}, \citenamefont {Murgia},\ and\ \citenamefont
  {Zaccheddu}}]{DAlesio:2020wjq}%
  \BibitemOpen
  \bibfield  {author} {\bibinfo {author} {\bibfnamefont {U.}~\bibnamefont
  {D'Alesio}}, \bibinfo {author} {\bibfnamefont {F.}~\bibnamefont {Murgia}}, \
  and\ \bibinfo {author} {\bibfnamefont {M.}~\bibnamefont {Zaccheddu}},\ }\href
  {\doibase 10.1103/PhysRevD.102.054001} {\bibfield  {journal} {\bibinfo
  {journal} {Phys. Rev. D}\ }\textbf {\bibinfo {volume} {102}},\ \bibinfo
  {pages} {054001} (\bibinfo {year} {2020})},\ \Eprint
  {http://arxiv.org/abs/2003.01128} {arXiv:2003.01128 [hep-ph]} \BibitemShut
  {NoStop}%
\bibitem [{\citenamefont {Callos}\ \emph {et~al.}(2020)\citenamefont {Callos},
  \citenamefont {Kang},\ and\ \citenamefont {Terry}}]{Callos:2020qtu}%
  \BibitemOpen
  \bibfield  {author} {\bibinfo {author} {\bibfnamefont {D.}~\bibnamefont
  {Callos}}, \bibinfo {author} {\bibfnamefont {Z.-B.}\ \bibnamefont {Kang}}, \
  and\ \bibinfo {author} {\bibfnamefont {J.}~\bibnamefont {Terry}},\ }\href
  {\doibase 10.1103/PhysRevD.102.096007} {\bibfield  {journal} {\bibinfo
  {journal} {Phys. Rev. D}\ }\textbf {\bibinfo {volume} {102}},\ \bibinfo
  {pages} {096007} (\bibinfo {year} {2020})},\ \Eprint
  {http://arxiv.org/abs/2003.04828} {arXiv:2003.04828 [hep-ph]} \BibitemShut
  {NoStop}%
\bibitem [{\citenamefont {Chen}\ \emph {et~al.}(2021)\citenamefont {Chen},
  \citenamefont {Liang}, \citenamefont {Pan}, \citenamefont {Song},\ and\
  \citenamefont {Wei}}]{Chen:2021hdn}%
  \BibitemOpen
  \bibfield  {author} {\bibinfo {author} {\bibfnamefont {K.-b.}\ \bibnamefont
  {Chen}}, \bibinfo {author} {\bibfnamefont {Z.-t.}\ \bibnamefont {Liang}},
  \bibinfo {author} {\bibfnamefont {Y.-l.}\ \bibnamefont {Pan}}, \bibinfo
  {author} {\bibfnamefont {Y.-k.}\ \bibnamefont {Song}}, \ and\ \bibinfo
  {author} {\bibfnamefont {S.-y.}\ \bibnamefont {Wei}},\ }\href@noop {} {\
  (\bibinfo {year} {2021})},\ \Eprint {http://arxiv.org/abs/2102.00658}
  {arXiv:2102.00658 [hep-ph]} \BibitemShut {NoStop}%
\bibitem [{\citenamefont {Collins}\ and\ \citenamefont
  {Soper}(1982{\natexlab{a}})}]{Collins:1981uw}%
  \BibitemOpen
  \bibfield  {author} {\bibinfo {author} {\bibfnamefont {J.~C.}\ \bibnamefont
  {Collins}}\ and\ \bibinfo {author} {\bibfnamefont {D.~E.}\ \bibnamefont
  {Soper}},\ }\href {\doibase 10.1016/0550-3213(82)90021-9} {\bibfield
  {journal} {\bibinfo  {journal} {Nucl. Phys.}\ }\textbf {\bibinfo {volume}
  {B194}},\ \bibinfo {pages} {445} (\bibinfo {year}
  {1982}{\natexlab{a}})}\BibitemShut {NoStop}%
\bibitem [{\citenamefont {Collins}\ and\ \citenamefont
  {Soper}(1982{\natexlab{b}})}]{Collins:1981va}%
  \BibitemOpen
  \bibfield  {author} {\bibinfo {author} {\bibfnamefont {J.~C.}\ \bibnamefont
  {Collins}}\ and\ \bibinfo {author} {\bibfnamefont {D.~E.}\ \bibnamefont
  {Soper}},\ }\href {\doibase 10.1016/0550-3213(82)90453-9} {\bibfield
  {journal} {\bibinfo  {journal} {Nucl.Phys.}\ }\textbf {\bibinfo {volume}
  {B197}},\ \bibinfo {pages} {446} (\bibinfo {year}
  {1982}{\natexlab{b}})}\BibitemShut {NoStop}%
\bibitem [{\citenamefont {Kang}\ \emph {et~al.}(2020)\citenamefont {Kang},
  \citenamefont {Shao},\ and\ \citenamefont {Zhao}}]{Kang:2020yqw}%
  \BibitemOpen
  \bibfield  {author} {\bibinfo {author} {\bibfnamefont {Z.-B.}\ \bibnamefont
  {Kang}}, \bibinfo {author} {\bibfnamefont {D.~Y.}\ \bibnamefont {Shao}}, \
  and\ \bibinfo {author} {\bibfnamefont {F.}~\bibnamefont {Zhao}},\ }\href
  {\doibase 10.1007/JHEP12(2020)127} {\bibfield  {journal} {\bibinfo  {journal}
  {JHEP}\ }\textbf {\bibinfo {volume} {12}},\ \bibinfo {pages} {127} (\bibinfo
  {year} {2020})},\ \Eprint {http://arxiv.org/abs/2007.14425} {arXiv:2007.14425
  [hep-ph]} \BibitemShut {NoStop}%
\bibitem [{\citenamefont {Boglione}\ and\ \citenamefont
  {Simonelli}(2020)}]{Boglione:2020auc}%
  \BibitemOpen
  \bibfield  {author} {\bibinfo {author} {\bibfnamefont {M.}~\bibnamefont
  {Boglione}}\ and\ \bibinfo {author} {\bibfnamefont {A.}~\bibnamefont
  {Simonelli}},\ }\href@noop {} {\  (\bibinfo {year} {2020})},\ \Eprint
  {http://arxiv.org/abs/2011.07366} {arXiv:2011.07366 [hep-ph]} \BibitemShut
  {NoStop}%
\bibitem [{\citenamefont {Makris}\ \emph {et~al.}(2020)\citenamefont {Makris},
  \citenamefont {Ringer},\ and\ \citenamefont {Waalewijn}}]{Makris:2020ltr}%
  \BibitemOpen
  \bibfield  {author} {\bibinfo {author} {\bibfnamefont {Y.}~\bibnamefont
  {Makris}}, \bibinfo {author} {\bibfnamefont {F.}~\bibnamefont {Ringer}}, \
  and\ \bibinfo {author} {\bibfnamefont {W.~J.}\ \bibnamefont {Waalewijn}},\
  }\href@noop {} {\  (\bibinfo {year} {2020})},\ \Eprint
  {http://arxiv.org/abs/2009.11871} {arXiv:2009.11871 [hep-ph]} \BibitemShut
  {NoStop}%
\bibitem [{\citenamefont {Qiu}\ and\ \citenamefont
  {Sterman}(1999)}]{Qiu:1998ia}%
  \BibitemOpen
  \bibfield  {author} {\bibinfo {author} {\bibfnamefont {J.-w.}\ \bibnamefont
  {Qiu}}\ and\ \bibinfo {author} {\bibfnamefont {G.~F.}\ \bibnamefont
  {Sterman}},\ }\href {\doibase 10.1103/PhysRevD.59.014004} {\bibfield
  {journal} {\bibinfo  {journal} {Phys. Rev. D}\ }\textbf {\bibinfo {volume}
  {59}},\ \bibinfo {pages} {014004} (\bibinfo {year} {1999})},\ \Eprint
  {http://arxiv.org/abs/hep-ph/9806356} {arXiv:hep-ph/9806356} \BibitemShut
  {NoStop}%
\bibitem [{\citenamefont {Metz}\ and\ \citenamefont
  {Pitonyak}(2013)}]{Metz:2012ct}%
  \BibitemOpen
  \bibfield  {author} {\bibinfo {author} {\bibfnamefont {A.}~\bibnamefont
  {Metz}}\ and\ \bibinfo {author} {\bibfnamefont {D.}~\bibnamefont
  {Pitonyak}},\ }\href {\doibase 10.1016/j.physletb.2013.05.043} {\bibfield
  {journal} {\bibinfo  {journal} {Phys. Lett. B}\ }\textbf {\bibinfo {volume}
  {723}},\ \bibinfo {pages} {365} (\bibinfo {year} {2013})},\ \bibinfo {note}
  {[Erratum: Phys.Lett.B 762, 549--549 (2016)]},\ \Eprint
  {http://arxiv.org/abs/1212.5037} {arXiv:1212.5037 [hep-ph]} \BibitemShut
  {NoStop}%
\bibitem [{\citenamefont {Lu}(1995)}]{Lu:1995rp}%
  \BibitemOpen
  \bibfield  {author} {\bibinfo {author} {\bibfnamefont {W.}~\bibnamefont
  {Lu}},\ }\href {\doibase 10.1103/PhysRevD.51.5305} {\bibfield  {journal}
  {\bibinfo  {journal} {Phys. Rev. D}\ }\textbf {\bibinfo {volume} {51}},\
  \bibinfo {pages} {5305} (\bibinfo {year} {1995})},\ \Eprint
  {http://arxiv.org/abs/hep-ph/9505361} {arXiv:hep-ph/9505361} \BibitemShut
  {NoStop}%
\bibitem [{\citenamefont {De~Rujula}\ \emph {et~al.}(1971)\citenamefont
  {De~Rujula}, \citenamefont {Kaplan},\ and\ \citenamefont
  {De~Rafael}}]{DeRujula:1972te}%
  \BibitemOpen
  \bibfield  {author} {\bibinfo {author} {\bibfnamefont {A.}~\bibnamefont
  {De~Rujula}}, \bibinfo {author} {\bibfnamefont {J.}~\bibnamefont {Kaplan}}, \
  and\ \bibinfo {author} {\bibfnamefont {E.}~\bibnamefont {De~Rafael}},\ }\href
  {\doibase 10.1016/0550-3213(71)90460-3} {\bibfield  {journal} {\bibinfo
  {journal} {Nucl. Phys. B}\ }\textbf {\bibinfo {volume} {35}},\ \bibinfo
  {pages} {365} (\bibinfo {year} {1971})}\BibitemShut {NoStop}%
\bibitem [{\citenamefont {Goeke}\ \emph {et~al.}(2005)\citenamefont {Goeke},
  \citenamefont {Metz},\ and\ \citenamefont {Schlegel}}]{Goeke:2005hb}%
  \BibitemOpen
  \bibfield  {author} {\bibinfo {author} {\bibfnamefont {K.}~\bibnamefont
  {Goeke}}, \bibinfo {author} {\bibfnamefont {A.}~\bibnamefont {Metz}}, \ and\
  \bibinfo {author} {\bibfnamefont {M.}~\bibnamefont {Schlegel}},\ }\href@noop
  {} {\bibfield  {journal} {\bibinfo  {journal} {Phys. Lett.}\ }\textbf
  {\bibinfo {volume} {B618}},\ \bibinfo {pages} {90} (\bibinfo {year}
  {2005})},\ \Eprint {http://arxiv.org/abs/hep-ph/0504130} {hep-ph/0504130}
  \BibitemShut {NoStop}%
\bibitem [{\citenamefont {Metz}\ and\ \citenamefont
  {Vossen}(2016)}]{Metz:2016swz}%
  \BibitemOpen
  \bibfield  {author} {\bibinfo {author} {\bibfnamefont {A.}~\bibnamefont
  {Metz}}\ and\ \bibinfo {author} {\bibfnamefont {A.}~\bibnamefont {Vossen}},\
  }\href {\doibase 10.1016/j.ppnp.2016.08.003} {\bibfield  {journal} {\bibinfo
  {journal} {Prog. Part. Nucl. Phys.}\ }\textbf {\bibinfo {volume} {91}},\
  \bibinfo {pages} {136} (\bibinfo {year} {2016})},\ \Eprint
  {http://arxiv.org/abs/1607.02521} {arXiv:1607.02521 [hep-ex]} \BibitemShut
  {NoStop}%
\bibitem [{\citenamefont {Christ}\ and\ \citenamefont
  {Lee}(1966)}]{Christ:1966zz}%
  \BibitemOpen
  \bibfield  {author} {\bibinfo {author} {\bibfnamefont {N.}~\bibnamefont
  {Christ}}\ and\ \bibinfo {author} {\bibfnamefont {T.}~\bibnamefont {Lee}},\
  }\href {\doibase 10.1103/PhysRev.143.1310} {\bibfield  {journal} {\bibinfo
  {journal} {Phys. Rev.}\ }\textbf {\bibinfo {volume} {143}},\ \bibinfo {pages}
  {1310} (\bibinfo {year} {1966})}\BibitemShut {NoStop}%
\bibitem [{\citenamefont {Chiu}\ \emph
  {et~al.}(2012{\natexlab{a}})\citenamefont {Chiu}, \citenamefont {Jain},
  \citenamefont {Neill},\ and\ \citenamefont {Rothstein}}]{Chiu:2011qc}%
  \BibitemOpen
  \bibfield  {author} {\bibinfo {author} {\bibfnamefont {J.-y.}\ \bibnamefont
  {Chiu}}, \bibinfo {author} {\bibfnamefont {A.}~\bibnamefont {Jain}}, \bibinfo
  {author} {\bibfnamefont {D.}~\bibnamefont {Neill}}, \ and\ \bibinfo {author}
  {\bibfnamefont {I.~Z.}\ \bibnamefont {Rothstein}},\ }\href {\doibase
  10.1103/PhysRevLett.108.151601} {\bibfield  {journal} {\bibinfo  {journal}
  {Phys. Rev. Lett.}\ }\textbf {\bibinfo {volume} {108}},\ \bibinfo {pages}
  {151601} (\bibinfo {year} {2012}{\natexlab{a}})},\ \Eprint
  {http://arxiv.org/abs/1104.0881} {arXiv:1104.0881 [hep-ph]} \BibitemShut
  {NoStop}%
\bibitem [{\citenamefont {Chiu}\ \emph
  {et~al.}(2012{\natexlab{b}})\citenamefont {Chiu}, \citenamefont {Jain},
  \citenamefont {Neill},\ and\ \citenamefont {Rothstein}}]{Chiu:2012ir}%
  \BibitemOpen
  \bibfield  {author} {\bibinfo {author} {\bibfnamefont {J.-Y.}\ \bibnamefont
  {Chiu}}, \bibinfo {author} {\bibfnamefont {A.}~\bibnamefont {Jain}}, \bibinfo
  {author} {\bibfnamefont {D.}~\bibnamefont {Neill}}, \ and\ \bibinfo {author}
  {\bibfnamefont {I.~Z.}\ \bibnamefont {Rothstein}},\ }\href {\doibase
  10.1007/JHEP05(2012)084} {\bibfield  {journal} {\bibinfo  {journal} {JHEP}\
  }\textbf {\bibinfo {volume} {05}},\ \bibinfo {pages} {084} (\bibinfo {year}
  {2012}{\natexlab{b}})},\ \Eprint {http://arxiv.org/abs/1202.0814}
  {arXiv:1202.0814 [hep-ph]} \BibitemShut {NoStop}%
\bibitem [{\citenamefont {Ebert}\ \emph {et~al.}(2019)\citenamefont {Ebert},
  \citenamefont {Stewart},\ and\ \citenamefont {Zhao}}]{Ebert:2019okf}%
  \BibitemOpen
  \bibfield  {author} {\bibinfo {author} {\bibfnamefont {M.~A.}\ \bibnamefont
  {Ebert}}, \bibinfo {author} {\bibfnamefont {I.~W.}\ \bibnamefont {Stewart}},
  \ and\ \bibinfo {author} {\bibfnamefont {Y.}~\bibnamefont {Zhao}},\ }\href
  {\doibase 10.1007/JHEP09(2019)037} {\bibfield  {journal} {\bibinfo  {journal}
  {JHEP}\ }\textbf {\bibinfo {volume} {09}},\ \bibinfo {pages} {037} (\bibinfo
  {year} {2019})},\ \Eprint {http://arxiv.org/abs/1901.03685} {arXiv:1901.03685
  [hep-ph]} \BibitemShut {NoStop}%
\bibitem [{\citenamefont {Boer}\ \emph {et~al.}(2011)\citenamefont {Boer},
  \citenamefont {Gamberg}, \citenamefont {Musch},\ and\ \citenamefont
  {Prokudin}}]{Boer:2011xd}%
  \BibitemOpen
  \bibfield  {author} {\bibinfo {author} {\bibfnamefont {D.}~\bibnamefont
  {Boer}}, \bibinfo {author} {\bibfnamefont {L.}~\bibnamefont {Gamberg}},
  \bibinfo {author} {\bibfnamefont {B.}~\bibnamefont {Musch}}, \ and\ \bibinfo
  {author} {\bibfnamefont {A.}~\bibnamefont {Prokudin}},\ }\href {\doibase
  10.1007/JHEP10(2011)021} {\bibfield  {journal} {\bibinfo  {journal} {JHEP}\
  }\textbf {\bibinfo {volume} {1110}},\ \bibinfo {pages} {021} (\bibinfo {year}
  {2011})},\ \Eprint {http://arxiv.org/abs/1107.5294} {arXiv:1107.5294
  [hep-ph]} \BibitemShut {NoStop}%
\bibitem [{\citenamefont {Ji}\ \emph {et~al.}(2005)\citenamefont {Ji},
  \citenamefont {Ma},\ and\ \citenamefont {Yuan}}]{Ji:2004wu}%
  \BibitemOpen
  \bibfield  {author} {\bibinfo {author} {\bibfnamefont {X.-d.}\ \bibnamefont
  {Ji}}, \bibinfo {author} {\bibfnamefont {J.-p.}\ \bibnamefont {Ma}}, \ and\
  \bibinfo {author} {\bibfnamefont {F.}~\bibnamefont {Yuan}},\ }\href {\doibase
  10.1103/PhysRevD.71.034005} {\bibfield  {journal} {\bibinfo  {journal} {Phys.
  Rev. D}\ }\textbf {\bibinfo {volume} {71}},\ \bibinfo {pages} {034005}
  (\bibinfo {year} {2005})},\ \Eprint {http://arxiv.org/abs/hep-ph/0404183}
  {arXiv:hep-ph/0404183} \BibitemShut {NoStop}%
\bibitem [{\citenamefont {Echevarria}\ \emph {et~al.}(2016)\citenamefont
  {Echevarria}, \citenamefont {Scimemi},\ and\ \citenamefont
  {Vladimirov}}]{Echevarria:2015byo}%
  \BibitemOpen
  \bibfield  {author} {\bibinfo {author} {\bibfnamefont {M.~G.}\ \bibnamefont
  {Echevarria}}, \bibinfo {author} {\bibfnamefont {I.}~\bibnamefont {Scimemi}},
  \ and\ \bibinfo {author} {\bibfnamefont {A.}~\bibnamefont {Vladimirov}},\
  }\href {\doibase 10.1103/PhysRevD.93.054004} {\bibfield  {journal} {\bibinfo
  {journal} {Phys. Rev. D}\ }\textbf {\bibinfo {volume} {93}},\ \bibinfo
  {pages} {054004} (\bibinfo {year} {2016})},\ \Eprint
  {http://arxiv.org/abs/1511.05590} {arXiv:1511.05590 [hep-ph]} \BibitemShut
  {NoStop}%
\bibitem [{\citenamefont {Dasgupta}\ and\ \citenamefont
  {Salam}(2001)}]{Dasgupta:2001sh}%
  \BibitemOpen
  \bibfield  {author} {\bibinfo {author} {\bibfnamefont {M.}~\bibnamefont
  {Dasgupta}}\ and\ \bibinfo {author} {\bibfnamefont {G.~P.}\ \bibnamefont
  {Salam}},\ }\href {\doibase 10.1016/S0370-2693(01)00725-0} {\bibfield
  {journal} {\bibinfo  {journal} {Phys. Lett.}\ }\textbf {\bibinfo {volume}
  {B512}},\ \bibinfo {pages} {323} (\bibinfo {year} {2001})},\ \Eprint
  {http://arxiv.org/abs/hep-ph/0104277} {arXiv:hep-ph/0104277 [hep-ph]}
  \BibitemShut {NoStop}%
\bibitem [{\citenamefont {Becher}\ \emph
  {et~al.}(2016{\natexlab{a}})\citenamefont {Becher}, \citenamefont {Neubert},
  \citenamefont {Rothen},\ and\ \citenamefont {Shao}}]{Becher:2015hka}%
  \BibitemOpen
  \bibfield  {author} {\bibinfo {author} {\bibfnamefont {T.}~\bibnamefont
  {Becher}}, \bibinfo {author} {\bibfnamefont {M.}~\bibnamefont {Neubert}},
  \bibinfo {author} {\bibfnamefont {L.}~\bibnamefont {Rothen}}, \ and\ \bibinfo
  {author} {\bibfnamefont {D.~Y.}\ \bibnamefont {Shao}},\ }\href {\doibase
  10.1103/PhysRevLett.116.192001} {\bibfield  {journal} {\bibinfo  {journal}
  {Phys. Rev. Lett.}\ }\textbf {\bibinfo {volume} {116}},\ \bibinfo {pages}
  {192001} (\bibinfo {year} {2016}{\natexlab{a}})},\ \Eprint
  {http://arxiv.org/abs/1508.06645} {arXiv:1508.06645 [hep-ph]} \BibitemShut
  {NoStop}%
\bibitem [{\citenamefont {Becher}\ \emph
  {et~al.}(2016{\natexlab{b}})\citenamefont {Becher}, \citenamefont {Neubert},
  \citenamefont {Rothen},\ and\ \citenamefont {Shao}}]{Becher:2016mmh}%
  \BibitemOpen
  \bibfield  {author} {\bibinfo {author} {\bibfnamefont {T.}~\bibnamefont
  {Becher}}, \bibinfo {author} {\bibfnamefont {M.}~\bibnamefont {Neubert}},
  \bibinfo {author} {\bibfnamefont {L.}~\bibnamefont {Rothen}}, \ and\ \bibinfo
  {author} {\bibfnamefont {D.~Y.}\ \bibnamefont {Shao}},\ }\href {\doibase
  10.1007/JHEP11(2016)019, 10.1007/JHEP05(2017)154} {\bibfield  {journal}
  {\bibinfo  {journal} {JHEP}\ }\textbf {\bibinfo {volume} {11}},\ \bibinfo
  {pages} {019} (\bibinfo {year} {2016}{\natexlab{b}})},\ \bibinfo {note}
  {[Erratum: JHEP05,154(2017)]},\ \Eprint {http://arxiv.org/abs/1605.02737}
  {arXiv:1605.02737 [hep-ph]} \BibitemShut {NoStop}%
\bibitem [{\citenamefont {Becher}\ \emph
  {et~al.}(2016{\natexlab{c}})\citenamefont {Becher}, \citenamefont {Pecjak},\
  and\ \citenamefont {Shao}}]{Becher:2016omr}%
  \BibitemOpen
  \bibfield  {author} {\bibinfo {author} {\bibfnamefont {T.}~\bibnamefont
  {Becher}}, \bibinfo {author} {\bibfnamefont {B.~D.}\ \bibnamefont {Pecjak}},
  \ and\ \bibinfo {author} {\bibfnamefont {D.~Y.}\ \bibnamefont {Shao}},\
  }\href {\doibase 10.1007/JHEP12(2016)018} {\bibfield  {journal} {\bibinfo
  {journal} {JHEP}\ }\textbf {\bibinfo {volume} {12}},\ \bibinfo {pages} {018}
  (\bibinfo {year} {2016}{\natexlab{c}})},\ \Eprint
  {http://arxiv.org/abs/1610.01608} {arXiv:1610.01608 [hep-ph]} \BibitemShut
  {NoStop}%
\bibitem [{\citenamefont {Becher}\ \emph {et~al.}(2017)\citenamefont {Becher},
  \citenamefont {Rahn},\ and\ \citenamefont {Shao}}]{Becher:2017nof}%
  \BibitemOpen
  \bibfield  {author} {\bibinfo {author} {\bibfnamefont {T.}~\bibnamefont
  {Becher}}, \bibinfo {author} {\bibfnamefont {R.}~\bibnamefont {Rahn}}, \ and\
  \bibinfo {author} {\bibfnamefont {D.~Y.}\ \bibnamefont {Shao}},\ }\href
  {\doibase 10.1007/JHEP10(2017)030} {\bibfield  {journal} {\bibinfo  {journal}
  {JHEP}\ }\textbf {\bibinfo {volume} {10}},\ \bibinfo {pages} {030} (\bibinfo
  {year} {2017})},\ \Eprint {http://arxiv.org/abs/1708.04516} {arXiv:1708.04516
  [hep-ph]} \BibitemShut {NoStop}%
\bibitem [{\citenamefont {Caron-Huot}(2018)}]{Caron-Huot:2015bja}%
  \BibitemOpen
  \bibfield  {author} {\bibinfo {author} {\bibfnamefont {S.}~\bibnamefont
  {Caron-Huot}},\ }\href {\doibase 10.1007/JHEP03(2018)036} {\bibfield
  {journal} {\bibinfo  {journal} {JHEP}\ }\textbf {\bibinfo {volume} {03}},\
  \bibinfo {pages} {036} (\bibinfo {year} {2018})},\ \Eprint
  {http://arxiv.org/abs/1501.03754} {arXiv:1501.03754 [hep-ph]} \BibitemShut
  {NoStop}%
\bibitem [{\citenamefont {Nagy}\ and\ \citenamefont
  {Soper}(2016)}]{Nagy:2016pwq}%
  \BibitemOpen
  \bibfield  {author} {\bibinfo {author} {\bibfnamefont {Z.}~\bibnamefont
  {Nagy}}\ and\ \bibinfo {author} {\bibfnamefont {D.~E.}\ \bibnamefont
  {Soper}},\ }\href {\doibase 10.1007/JHEP10(2016)019} {\bibfield  {journal}
  {\bibinfo  {journal} {JHEP}\ }\textbf {\bibinfo {volume} {10}},\ \bibinfo
  {pages} {019} (\bibinfo {year} {2016})},\ \Eprint
  {http://arxiv.org/abs/1605.05845} {arXiv:1605.05845 [hep-ph]} \BibitemShut
  {NoStop}%
\bibitem [{\citenamefont {Nagy}\ and\ \citenamefont
  {Soper}(2018)}]{Nagy:2017ggp}%
  \BibitemOpen
  \bibfield  {author} {\bibinfo {author} {\bibfnamefont {Z.}~\bibnamefont
  {Nagy}}\ and\ \bibinfo {author} {\bibfnamefont {D.~E.}\ \bibnamefont
  {Soper}},\ }\href {\doibase 10.1103/PhysRevD.98.014034} {\bibfield  {journal}
  {\bibinfo  {journal} {Phys. Rev. D}\ }\textbf {\bibinfo {volume} {98}},\
  \bibinfo {pages} {014034} (\bibinfo {year} {2018})},\ \Eprint
  {http://arxiv.org/abs/1705.08093} {arXiv:1705.08093 [hep-ph]} \BibitemShut
  {NoStop}%
\bibitem [{\citenamefont {Collins}\ \emph {et~al.}(1985)\citenamefont
  {Collins}, \citenamefont {Soper},\ and\ \citenamefont
  {Sterman}}]{Collins:1984kg}%
  \BibitemOpen
  \bibfield  {author} {\bibinfo {author} {\bibfnamefont {J.~C.}\ \bibnamefont
  {Collins}}, \bibinfo {author} {\bibfnamefont {D.~E.}\ \bibnamefont {Soper}},
  \ and\ \bibinfo {author} {\bibfnamefont {G.~F.}\ \bibnamefont {Sterman}},\
  }\href {\doibase 10.1016/0550-3213(85)90479-1} {\bibfield  {journal}
  {\bibinfo  {journal} {Nucl. Phys.}\ }\textbf {\bibinfo {volume} {B250}},\
  \bibinfo {pages} {199} (\bibinfo {year} {1985})}\BibitemShut {NoStop}%
\bibitem [{\citenamefont {Bacchetta}\ \emph {et~al.}(2004)\citenamefont
  {Bacchetta}, \citenamefont {D'Alesio}, \citenamefont {Diehl},\ and\
  \citenamefont {Miller}}]{Bacchetta:2004jz}%
  \BibitemOpen
  \bibfield  {author} {\bibinfo {author} {\bibfnamefont {A.}~\bibnamefont
  {Bacchetta}}, \bibinfo {author} {\bibfnamefont {U.}~\bibnamefont {D'Alesio}},
  \bibinfo {author} {\bibfnamefont {M.}~\bibnamefont {Diehl}}, \ and\ \bibinfo
  {author} {\bibfnamefont {C.}~\bibnamefont {Miller}},\ }\href {\doibase
  10.1103/PhysRevD.70.117504} {\bibfield  {journal} {\bibinfo  {journal} {Phys.
  Rev. D}\ }\textbf {\bibinfo {volume} {70}},\ \bibinfo {pages} {117504}
  (\bibinfo {year} {2004})},\ \Eprint {http://arxiv.org/abs/hep-ph/0410050}
  {arXiv:hep-ph/0410050} \BibitemShut {NoStop}%
\bibitem [{\citenamefont {Boer}\ \emph {et~al.}(1998)\citenamefont {Boer},
  \citenamefont {Jakob},\ and\ \citenamefont {Mulders}}]{Boer:1997qn}%
  \BibitemOpen
  \bibfield  {author} {\bibinfo {author} {\bibfnamefont {D.}~\bibnamefont
  {Boer}}, \bibinfo {author} {\bibfnamefont {R.}~\bibnamefont {Jakob}}, \ and\
  \bibinfo {author} {\bibfnamefont {P.~J.}\ \bibnamefont {Mulders}},\ }\href
  {\doibase 10.1016/S0370-2693(98)00136-1} {\bibfield  {journal} {\bibinfo
  {journal} {Phys. Lett.}\ }\textbf {\bibinfo {volume} {B424}},\ \bibinfo
  {pages} {143} (\bibinfo {year} {1998})},\ \Eprint
  {http://arxiv.org/abs/hep-ph/9711488} {arXiv:hep-ph/9711488} \BibitemShut
  {NoStop}%
\bibitem [{\citenamefont {Collins}\ \emph {et~al.}(1989)\citenamefont
  {Collins}, \citenamefont {Soper},\ and\ \citenamefont
  {Sterman}}]{Collins:1989gx}%
  \BibitemOpen
  \bibfield  {author} {\bibinfo {author} {\bibfnamefont {J.~C.}\ \bibnamefont
  {Collins}}, \bibinfo {author} {\bibfnamefont {D.~E.}\ \bibnamefont {Soper}},
  \ and\ \bibinfo {author} {\bibfnamefont {G.~F.}\ \bibnamefont {Sterman}},\
  }\href {\doibase 10.1142/9789814503266_0001} {\bibfield  {journal} {\bibinfo
  {journal} {Adv. Ser. Direct. High Energy Phys.}\ }\textbf {\bibinfo {volume}
  {5}},\ \bibinfo {pages} {1} (\bibinfo {year} {1989})},\ \Eprint
  {http://arxiv.org/abs/hep-ph/0409313} {arXiv:hep-ph/0409313} \BibitemShut
  {NoStop}%
\bibitem [{\citenamefont {Bauer}\ \emph {et~al.}(2002)\citenamefont {Bauer},
  \citenamefont {Fleming}, \citenamefont {Pirjol}, \citenamefont {Rothstein},\
  and\ \citenamefont {Stewart}}]{Bauer:2002nz}%
  \BibitemOpen
  \bibfield  {author} {\bibinfo {author} {\bibfnamefont {C.~W.}\ \bibnamefont
  {Bauer}}, \bibinfo {author} {\bibfnamefont {S.}~\bibnamefont {Fleming}},
  \bibinfo {author} {\bibfnamefont {D.}~\bibnamefont {Pirjol}}, \bibinfo
  {author} {\bibfnamefont {I.~Z.}\ \bibnamefont {Rothstein}}, \ and\ \bibinfo
  {author} {\bibfnamefont {I.~W.}\ \bibnamefont {Stewart}},\ }\href {\doibase
  10.1103/PhysRevD.66.014017} {\bibfield  {journal} {\bibinfo  {journal} {Phys.
  Rev. D}\ }\textbf {\bibinfo {volume} {66}},\ \bibinfo {pages} {014017}
  (\bibinfo {year} {2002})},\ \Eprint {http://arxiv.org/abs/hep-ph/0202088}
  {arXiv:hep-ph/0202088} \BibitemShut {NoStop}%
\bibitem [{\citenamefont {Gamberg}\ \emph {et~al.}(2006)\citenamefont
  {Gamberg}, \citenamefont {Hwang}, \citenamefont {Metz},\ and\ \citenamefont
  {Schlegel}}]{Gamberg:2006ru}%
  \BibitemOpen
  \bibfield  {author} {\bibinfo {author} {\bibfnamefont {L.~P.}\ \bibnamefont
  {Gamberg}}, \bibinfo {author} {\bibfnamefont {D.~S.}\ \bibnamefont {Hwang}},
  \bibinfo {author} {\bibfnamefont {A.}~\bibnamefont {Metz}}, \ and\ \bibinfo
  {author} {\bibfnamefont {M.}~\bibnamefont {Schlegel}},\ }\href@noop {}
  {\bibfield  {journal} {\bibinfo  {journal} {Phys. Lett.}\ }\textbf {\bibinfo
  {volume} {B639}},\ \bibinfo {pages} {508} (\bibinfo {year} {2006})},\ \Eprint
  {http://arxiv.org/abs/hep-ph/0604022} {hep-ph/0604022} \BibitemShut {NoStop}%
\bibitem [{\citenamefont {Bacchetta}\ \emph {et~al.}(2019)\citenamefont
  {Bacchetta}, \citenamefont {Bozzi}, \citenamefont {Echevarria}, \citenamefont
  {Pisano}, \citenamefont {Prokudin},\ and\ \citenamefont
  {Radici}}]{Bacchetta:2019qkv}%
  \BibitemOpen
  \bibfield  {author} {\bibinfo {author} {\bibfnamefont {A.}~\bibnamefont
  {Bacchetta}}, \bibinfo {author} {\bibfnamefont {G.}~\bibnamefont {Bozzi}},
  \bibinfo {author} {\bibfnamefont {M.~G.}\ \bibnamefont {Echevarria}},
  \bibinfo {author} {\bibfnamefont {C.}~\bibnamefont {Pisano}}, \bibinfo
  {author} {\bibfnamefont {A.}~\bibnamefont {Prokudin}}, \ and\ \bibinfo
  {author} {\bibfnamefont {M.}~\bibnamefont {Radici}},\ }\href {\doibase
  10.1016/j.physletb.2019.134850} {\bibfield  {journal} {\bibinfo  {journal}
  {Phys. Lett. B}\ }\textbf {\bibinfo {volume} {797}},\ \bibinfo {pages}
  {134850} (\bibinfo {year} {2019})},\ \Eprint
  {http://arxiv.org/abs/1906.07037} {arXiv:1906.07037 [hep-ph]} \BibitemShut
  {NoStop}%
\bibitem [{\citenamefont {Collins}(2011)}]{Collins:2011qcdbook}%
  \BibitemOpen
  \bibfield  {author} {\bibinfo {author} {\bibfnamefont {J.~C.}\ \bibnamefont
  {Collins}},\ }\href@noop {} {\emph {\bibinfo {title} {Foundations of
  Perturbative QCD}}}\ (\bibinfo  {publisher} {Cambridge University Press},\
  \bibinfo {address} {Cambridge},\ \bibinfo {year} {2011})\BibitemShut
  {NoStop}%
\bibitem [{\citenamefont {Collins}\ and\ \citenamefont
  {Rogers}(2015{\natexlab{a}})}]{Collins:2014jpa}%
  \BibitemOpen
  \bibfield  {author} {\bibinfo {author} {\bibfnamefont {J.}~\bibnamefont
  {Collins}}\ and\ \bibinfo {author} {\bibfnamefont {T.}~\bibnamefont
  {Rogers}},\ }\href {\doibase 10.1103/PhysRevD.91.074020} {\bibfield
  {journal} {\bibinfo  {journal} {Phys.Rev.}\ }\textbf {\bibinfo {volume}
  {D91}},\ \bibinfo {pages} {074020} (\bibinfo {year} {2015}{\natexlab{a}})},\
  \Eprint {http://arxiv.org/abs/1412.3820} {arXiv:1412.3820 [hep-ph]}
  \BibitemShut {NoStop}%
\bibitem [{\citenamefont {Konychev}\ and\ \citenamefont
  {Nadolsky}(2006)}]{Konychev:2005iy}%
  \BibitemOpen
  \bibfield  {author} {\bibinfo {author} {\bibfnamefont {A.~V.}\ \bibnamefont
  {Konychev}}\ and\ \bibinfo {author} {\bibfnamefont {P.~M.}\ \bibnamefont
  {Nadolsky}},\ }\href {\doibase 10.1016/j.physletb.2005.12.063} {\bibfield
  {journal} {\bibinfo  {journal} {Phys.Lett.}\ }\textbf {\bibinfo {volume}
  {B633}},\ \bibinfo {pages} {710} (\bibinfo {year} {2006})},\ \Eprint
  {http://arxiv.org/abs/hep-ph/0506225} {arXiv:hep-ph/0506225 [hep-ph]}
  \BibitemShut {NoStop}%
\bibitem [{\citenamefont {Collins}\ and\ \citenamefont
  {Rogers}(2015{\natexlab{b}})}]{Collins:2015dfa}%
  \BibitemOpen
  \bibfield  {author} {\bibinfo {author} {\bibfnamefont {J.}~\bibnamefont
  {Collins}}\ and\ \bibinfo {author} {\bibfnamefont {T.~C.}\ \bibnamefont
  {Rogers}},\ }\href {\doibase 10.22323/1.247.0189} {\bibfield  {journal}
  {\bibinfo  {journal} {PoS}\ }\textbf {\bibinfo {volume} {DIS2015}},\ \bibinfo
  {pages} {189} (\bibinfo {year} {2015}{\natexlab{b}})},\ \Eprint
  {http://arxiv.org/abs/1507.05542} {arXiv:1507.05542 [hep-ph]} \BibitemShut
  {NoStop}%
\bibitem [{\citenamefont {Albino}\ \emph {et~al.}(2008)\citenamefont {Albino},
  \citenamefont {Kniehl},\ and\ \citenamefont {Kramer}}]{Albino:2008fy}%
  \BibitemOpen
  \bibfield  {author} {\bibinfo {author} {\bibfnamefont {S.}~\bibnamefont
  {Albino}}, \bibinfo {author} {\bibfnamefont {B.}~\bibnamefont {Kniehl}}, \
  and\ \bibinfo {author} {\bibfnamefont {G.}~\bibnamefont {Kramer}},\ }\href
  {\doibase 10.1016/j.nuclphysb.2008.05.017} {\bibfield  {journal} {\bibinfo
  {journal} {Nucl. Phys. B}\ }\textbf {\bibinfo {volume} {803}},\ \bibinfo
  {pages} {42} (\bibinfo {year} {2008})},\ \Eprint
  {http://arxiv.org/abs/0803.2768} {arXiv:0803.2768 [hep-ph]} \BibitemShut
  {NoStop}%
\bibitem [{\citenamefont {Aidala}\ \emph {et~al.}(2014)\citenamefont {Aidala},
  \citenamefont {Field}, \citenamefont {Gamberg},\ and\ \citenamefont
  {Rogers}}]{Aidala:2014hva}%
  \BibitemOpen
  \bibfield  {author} {\bibinfo {author} {\bibfnamefont {C.~A.}\ \bibnamefont
  {Aidala}}, \bibinfo {author} {\bibfnamefont {B.}~\bibnamefont {Field}},
  \bibinfo {author} {\bibfnamefont {L.~P.}\ \bibnamefont {Gamberg}}, \ and\
  \bibinfo {author} {\bibfnamefont {T.~C.}\ \bibnamefont {Rogers}},\ }\href
  {\doibase 10.1103/PhysRevD.89.094002} {\bibfield  {journal} {\bibinfo
  {journal} {Phys. Rev. D}\ }\textbf {\bibinfo {volume} {89}},\ \bibinfo
  {pages} {094002} (\bibinfo {year} {2014})},\ \Eprint
  {http://arxiv.org/abs/1401.2654} {arXiv:1401.2654 [hep-ph]} \BibitemShut
  {NoStop}%
\bibitem [{\citenamefont {Sun}\ \emph {et~al.}(2018)\citenamefont {Sun},
  \citenamefont {Isaacson}, \citenamefont {Yuan},\ and\ \citenamefont
  {Yuan}}]{Su:2014wpa}%
  \BibitemOpen
  \bibfield  {author} {\bibinfo {author} {\bibfnamefont {P.}~\bibnamefont
  {Sun}}, \bibinfo {author} {\bibfnamefont {J.}~\bibnamefont {Isaacson}},
  \bibinfo {author} {\bibfnamefont {C.~P.}\ \bibnamefont {Yuan}}, \ and\
  \bibinfo {author} {\bibfnamefont {F.}~\bibnamefont {Yuan}},\ }\href {\doibase
  10.1142/S0217751X18410063} {\bibfield  {journal} {\bibinfo  {journal} {Int.
  J. Mod. Phys. A}\ }\textbf {\bibinfo {volume} {33}},\ \bibinfo {pages}
  {1841006} (\bibinfo {year} {2018})},\ \Eprint
  {http://arxiv.org/abs/1406.3073} {arXiv:1406.3073 [hep-ph]} \BibitemShut
  {NoStop}%
\bibitem [{\citenamefont {Kang}\ \emph {et~al.}(2021)\citenamefont {Kang},
  \citenamefont {Prokudin}, \citenamefont {Sato},\ and\ \citenamefont
  {Terry}}]{Kang:2019ctl}%
  \BibitemOpen
  \bibfield  {author} {\bibinfo {author} {\bibfnamefont {Z.-B.}\ \bibnamefont
  {Kang}}, \bibinfo {author} {\bibfnamefont {A.}~\bibnamefont {Prokudin}},
  \bibinfo {author} {\bibfnamefont {N.}~\bibnamefont {Sato}}, \ and\ \bibinfo
  {author} {\bibfnamefont {J.}~\bibnamefont {Terry}},\ }\href {\doibase
  10.1016/j.cpc.2020.107611} {\bibfield  {journal} {\bibinfo  {journal}
  {Comput. Phys. Commun.}\ }\textbf {\bibinfo {volume} {258}},\ \bibinfo
  {pages} {107611} (\bibinfo {year} {2021})},\ \Eprint
  {http://arxiv.org/abs/1906.05949} {arXiv:1906.05949 [hep-ph]} \BibitemShut
  {NoStop}%
\bibitem [{\citenamefont {Gamberg}\ \emph {et~al.}(2018)\citenamefont
  {Gamberg}, \citenamefont {Metz}, \citenamefont {Pitonyak},\ and\
  \citenamefont {Prokudin}}]{Gamberg:2017jha}%
  \BibitemOpen
  \bibfield  {author} {\bibinfo {author} {\bibfnamefont {L.}~\bibnamefont
  {Gamberg}}, \bibinfo {author} {\bibfnamefont {A.}~\bibnamefont {Metz}},
  \bibinfo {author} {\bibfnamefont {D.}~\bibnamefont {Pitonyak}}, \ and\
  \bibinfo {author} {\bibfnamefont {A.}~\bibnamefont {Prokudin}},\ }\href
  {\doibase 10.1016/j.physletb.2018.03.024} {\bibfield  {journal} {\bibinfo
  {journal} {Phys. Lett.}\ }\textbf {\bibinfo {volume} {B781}},\ \bibinfo
  {pages} {443} (\bibinfo {year} {2018})},\ \Eprint
  {http://arxiv.org/abs/1712.08116} {arXiv:1712.08116 [hep-ph]} \BibitemShut
  {NoStop}%
\bibitem [{\citenamefont {Kanazawa}\ \emph {et~al.}(2016)\citenamefont
  {Kanazawa}, \citenamefont {Koike}, \citenamefont {Metz}, \citenamefont
  {Pitonyak},\ and\ \citenamefont {Schlegel}}]{Kanazawa:2015ajw}%
  \BibitemOpen
  \bibfield  {author} {\bibinfo {author} {\bibfnamefont {K.}~\bibnamefont
  {Kanazawa}}, \bibinfo {author} {\bibfnamefont {Y.}~\bibnamefont {Koike}},
  \bibinfo {author} {\bibfnamefont {A.}~\bibnamefont {Metz}}, \bibinfo {author}
  {\bibfnamefont {D.}~\bibnamefont {Pitonyak}}, \ and\ \bibinfo {author}
  {\bibfnamefont {M.}~\bibnamefont {Schlegel}},\ }\href {\doibase
  10.1103/PhysRevD.93.054024} {\bibfield  {journal} {\bibinfo  {journal} {Phys.
  Rev.}\ }\textbf {\bibinfo {volume} {D93}},\ \bibinfo {pages} {054024}
  (\bibinfo {year} {2016})},\ \Eprint {http://arxiv.org/abs/1512.07233}
  {arXiv:1512.07233 [hep-ph]} \BibitemShut {NoStop}%
\bibitem [{\citenamefont {Collins}\ \emph {et~al.}(2016)\citenamefont
  {Collins}, \citenamefont {Gamberg}, \citenamefont {Prokudin}, \citenamefont
  {Rogers}, \citenamefont {Sato},\ and\ \citenamefont
  {Wang}}]{Collins:2016hqq}%
  \BibitemOpen
  \bibfield  {author} {\bibinfo {author} {\bibfnamefont {J.}~\bibnamefont
  {Collins}}, \bibinfo {author} {\bibfnamefont {L.}~\bibnamefont {Gamberg}},
  \bibinfo {author} {\bibfnamefont {A.}~\bibnamefont {Prokudin}}, \bibinfo
  {author} {\bibfnamefont {T.~C.}\ \bibnamefont {Rogers}}, \bibinfo {author}
  {\bibfnamefont {N.}~\bibnamefont {Sato}}, \ and\ \bibinfo {author}
  {\bibfnamefont {B.}~\bibnamefont {Wang}},\ }\href {\doibase
  10.1103/PhysRevD.94.034014} {\bibfield  {journal} {\bibinfo  {journal} {Phys.
  Rev.}\ }\textbf {\bibinfo {volume} {D94}},\ \bibinfo {pages} {034014}
  (\bibinfo {year} {2016})},\ \Eprint {http://arxiv.org/abs/1605.00671}
  {arXiv:1605.00671 [hep-ph]} \BibitemShut {NoStop}%
\bibitem [{\citenamefont {Qiu}\ \emph {et~al.}(2020)\citenamefont {Qiu},
  \citenamefont {Rogers},\ and\ \citenamefont {Wang}}]{Qiu:2020oqr}%
  \BibitemOpen
  \bibfield  {author} {\bibinfo {author} {\bibfnamefont {J.-W.}\ \bibnamefont
  {Qiu}}, \bibinfo {author} {\bibfnamefont {T.~C.}\ \bibnamefont {Rogers}}, \
  and\ \bibinfo {author} {\bibfnamefont {B.}~\bibnamefont {Wang}},\ }\href
  {\doibase 10.1103/PhysRevD.101.116017} {\bibfield  {journal} {\bibinfo
  {journal} {Phys. Rev. D}\ }\textbf {\bibinfo {volume} {101}},\ \bibinfo
  {pages} {116017} (\bibinfo {year} {2020})},\ \Eprint
  {http://arxiv.org/abs/2004.13193} {arXiv:2004.13193 [hep-ph]} \BibitemShut
  {NoStop}%
\end{thebibliography}%
\end{document}